\documentclass{article}

\usepackage{microtype}
\usepackage{graphicx}
\usepackage{subfigure}
\usepackage{booktabs} 
\usepackage{multirow}

\usepackage{hyperref}

\usepackage[accepted]{icml2024}

\usepackage{amsmath}
\usepackage{amssymb}
\usepackage{mathtools}
\usepackage{amsthm}

\usepackage[capitalize,noabbrev]{cleveref}

\theoremstyle{plain}
\newtheorem{theorem}{Theorem}[section]

\theoremstyle{definition}
\newtheorem{definition}[theorem]{Definition}

\theoremstyle{remark}

\usepackage[textsize=tiny]{todonotes}

\begin{document}

\twocolumn[
\icmltitle{A-PSRO: A Unified Strategy Learning Method with Advantage Function for Normal-form Games}



\icmlsetsymbol{equal}{*}

\begin{icmlauthorlist}
\icmlauthor{Yudong Hu}{yyy}
\icmlauthor{Haoran Li}{yyy}
\icmlauthor{Congying Han}{yyy}
\icmlauthor{Tiande Guo}{yyy}
\icmlauthor{Mingqiang Li}{yyy}
\icmlauthor{Bonan Li}{yyy}
\end{icmlauthorlist}

\icmlaffiliation{yyy}{University of Chinese Academy of Sciences}

\icmlcorrespondingauthor{Firstname1 Lastname1}{first1.last1@xxx.edu}
\icmlcorrespondingauthor{Firstname2 Lastname2}{first2.last2@www.uk}

\icmlkeywords{Machine Learning, ICML}

\vskip 0.3in
]




\begin{abstract}
Solving Nash equilibrium is the key challenge in normal-form games with large strategy spaces, where open-ended learning frameworks offer an efficient approach. In this work, we propose an innovative unified open-ended learning framework A-PSRO, \textit{i.e.}, \textbf{A}dvantage \textbf{P}olicy \textbf{S}pace \textbf{R}esponse \textbf{O}racle, as a comprehensive framework for both zero-sum and general-sum games. In particular, we introduce the advantage function as an enhanced evaluation metric for strategies, enabling a unified learning objective for agents engaged in normal-form games. We prove that the advantage function exhibits favorable properties and is connected with the Nash equilibrium, which can be used as an objective to guide agents to learn strategies efficiently. Our experiments reveal that A-PSRO achieves a considerable decrease in exploitability in zero-sum games and an escalation in rewards in general-sum games, significantly outperforming previous PSRO algorithms.
\end{abstract}

\section{Introduction}

Nash equilibrium in normal-form games (\textit{i.e.}, zero-sum games and general-sum games) is widely employed to model the strategic behavior of rational, utility-maximizing players in games. Recently, the field of Multiagent Reinforcement Learning (MARL) has made substantial progress in solving equilibrium in games, including AI in chess \& poker games \cite{7,14} and real-time strategy (RTS) games \cite{6,13}. In the context of non-cooperative multiagent games, the violation of the Markov property necessitates exploration of potential interactions with opponents for the purpose of strategy selection and decision making \cite{12}. For large-scale problems characterized by complex strategy spaces, the Policy Space Response Oracle (PSRO) offers an effective open-ended framework for problem-solving \cite{1}.

In zero-sum games, previous works have revealed the presence of transitive and cyclic structures \cite{5}. In this context, it has been established that diversity strategy approaches exhibit efficacy in learning Nash equilibrium within cyclic game scenarios \cite{11}. In the realm of modeling diversity within PSRO, several notable methods have been put forth, \textit{e.g.}, UDF-PSRO \cite{2} and UDM-PSRO \cite{3}. However, the objective of enhancing diversity does not guarantee the attainment of convergence towards Nash equilibrium. Our work introduces an original approach by incorporating the concept of advantage as an autonomous evaluative metric. We establish the mathematical equivalence between advantage maximization and Nash equilibrium, and provide a methodology for exploring strategies with higher advantage in the PSRO framework. Our theory demonstrates that A-PSRO can approach the Nash equilibrium deterministically, enabling agents to learn strategies efficiently.

In the context of general-sum games, the landscape is marked by the presence of multiple Nash equilibria, each associated with disparate rewards \cite{20}. This is different from zero-sum games, where the Minimax property allow any Nash equilibrium to be learning objectives for agents \cite{18}. Considering the predominant objective of individual agents to optimize their distinct rewards in general-sum games, the convergence of the system depends on consistency in the lookahead process of agents \cite{19}. When adopting an appropriate objective for the strategy learning of agents, the multiagent system will converge to the Nash equilibrium where the joint reward is maximized \cite{8}. In general-sum games, PSRO algorithms aiming at cooperation can learn the Nash equilibrium faster \cite{32}. However, the application of the PSRO algorithm to enhance the rewards of agents in general-sum games has seldom been considered in previous research. In this work, we illustrate the consistency of maximizing the advantage function with exploring equilibrium with optimal rewards. This enables A-PSRO to achieve higher rewards compared to other algorithms in the process of learning strategies.

To summarize, A-PSRO aims to maximize the advantage function during the agent's learning process, enhancing the efficiency of strategy learning across various game environments: 1) Symmetric zero-sum games. In symmetric zero-sum games, the advantage function exhibits favorable properties, such as Lipschitz continuity and convexity. These properties enable agents to deterministically approach the Nash equilibrium by exploring strategies with higher advantage. We introduce the advantage function as a modification to the existing approach based on diversity module, resulting in significant improvements in the learning of Nash equilibrium strategies. 2) Two-player general-sum games. The advantage function is non-convex in general-sum games, and its local maximum corresponds to Nash equilibria with different joint rewards. By exploring strategies near the global optimum with the objective of maximizing the advantage function, our algorithm converges to equilibrium with higher rewards. 3) Multi-player games. We extend the definition of the advantage function to multi-player games and provide a framework for the A-PSRO algorithm to learn strategies in such settings. We show experimentally that A-PSRO is also effective to learn Nash equilibrium strategies in multi-player games. 

We conduct experiments across diverse games to assess the performance of our algorithm. Results from these experiments show that agents using the A-PSRO algorithm acquire strategies with markedly reduced exploitability in zero-sum games. In general-sum games, the A-PSRO algorithm enables agents to learn strategies with global optimal rewards, preventing the algorithm from stagnating in a local optimal Nash equilibrium. These results underscore the effectiveness of the A-PSRO algorithm as a unified framework for normal-form games. We hope that this work will encourage further research in the field of PSRO.

\begin{figure*}[t]
\centering
\subfigure[Zero-sum game geometrical structure] {\label{fg3}
\includegraphics[width=0.9\columnwidth]{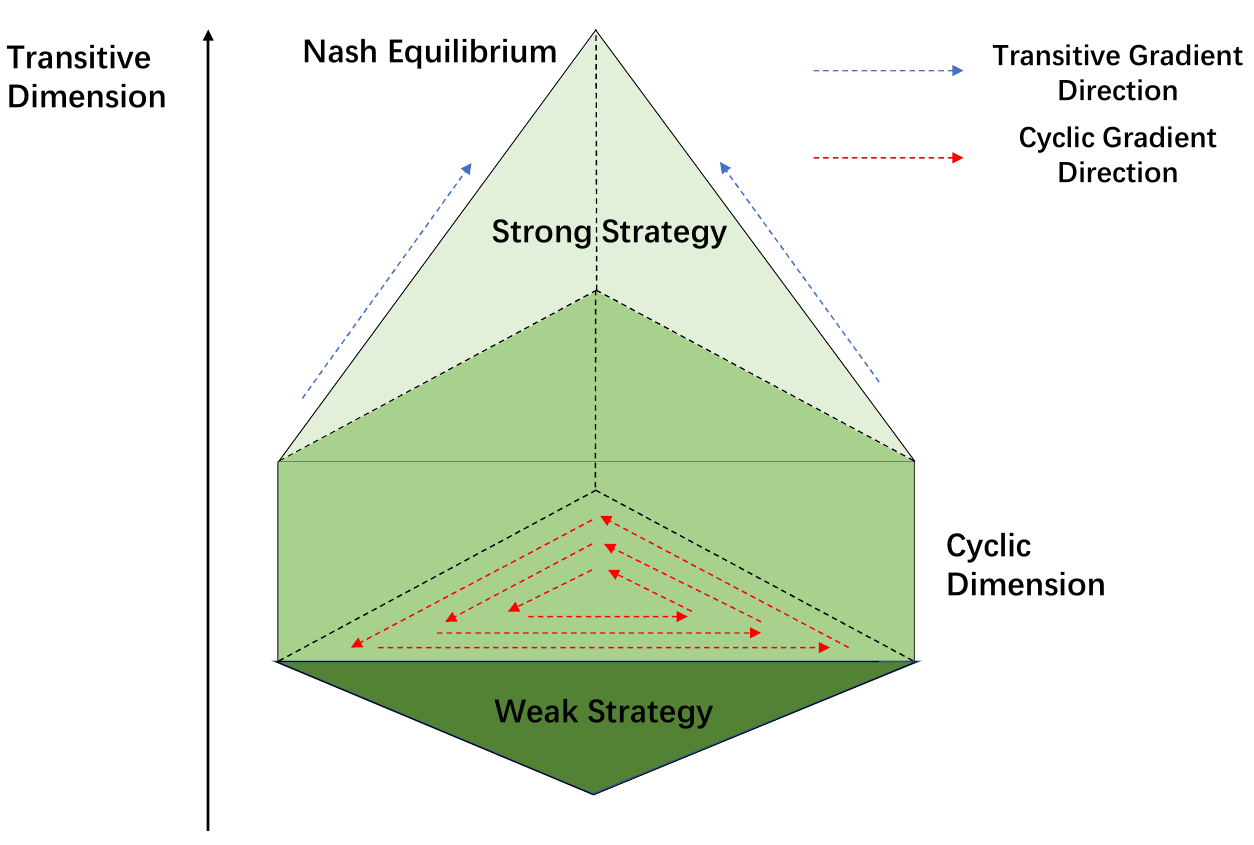}
}
\subfigure[General-sum game geometrical structure] {\label{fg2}
\includegraphics[width=0.9\columnwidth]{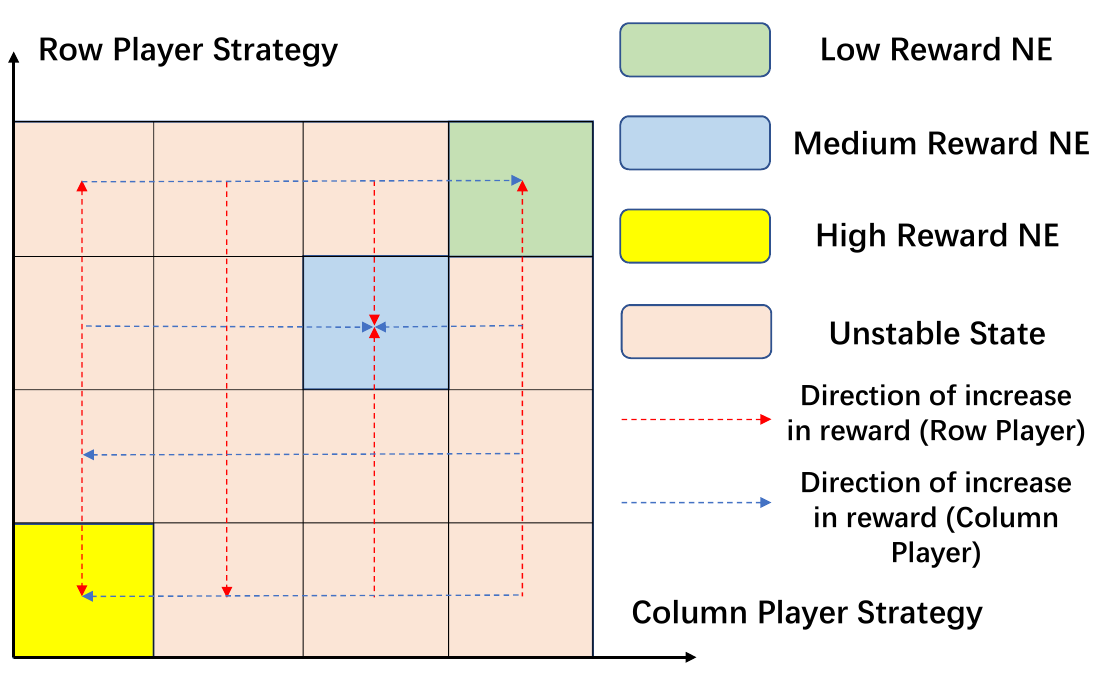}
}
\caption{The geometrical structure examples of zero-sum games and general-sum games. Figure (a) shows the structure of a zero-sum game with both transitive and cyclic dimensions. The direction of the strategy gradient refers to the expected updates for a strategy that maximize the reward. Figure (b) shows the structure of a general-sum game with multiple equilibia. The independent learning process of the agents leads to the update of the strategy in the direction indicated by the arrow.}
\label{fig2}
\vspace{-0.4cm}
\end{figure*}

\section{Notation and Background}

In this paper, we focus on normal-form games with finite dimension, typically represented by three key elements denoted as ($\mathcal{N},\mathcal{A},\mathcal{U}$). Here, $\mathcal{N}$ represents the players in the game, $\mathcal{A}$ denotes the action (pure strategy) space of the players, and $\mathcal{U}$ denotes the utility function or the payment matrix. In games, agents generally adopt strategies $\pi$ rather than actions $a \in \mathcal{A}$. $\pi$ is defined as a probability distribution over actions: $\pi = (p_1,p_2,\cdots,p_{\lvert \mathcal{A} \rvert}), \ \sum p_i =1$, where $p_i$ represents the probability of choosing action $a_i$. In this paper, we use $\pi^t_i$ to denote the $t$th strategy of player $i$.

Nash equilibrium (NE) characterizes a stable state in a game, where no agent can enhance its reward by unilaterally altering its strategy. For the joint strategy $(\pi_1, \cdots, \pi_{n})$, it is an NE when $\forall i \in \{1,\cdots,n\}$, $\pi_i$ is the best response (BR) to the strategies of other agents: $\forall \pi^*_i, \ U_i(\pi^*_i,\pi_{-i}) \leq U_i(\pi_i,\pi_{-i})$ ($\pi_{-i}$ represents the joint strategy except for agent $i$).

Exploitability is defined as the distance of joint strategy $(\pi_i,\pi_{-i})$ and the Nash Equilibrium:
\begin{equation}
    \mathcal{E}(\pi_i,\pi_{-i}) = \sum^n_{k=1} [\operatorname{max}_{\pi^*_k} U_k(\pi^*_k,\pi_{-k}) -U_k(\pi_k,\pi_{-k})]. 
\end{equation}
If the exploitability of a joint strategy $(\pi_i,\pi_{-i})$ is 0, it is a Nash equilibrium.

\subsection{Symmetric Zero-Sum Games with Transitive Dimension and Cyclic Dimension}
\label{Symmetric Zero-Sum Games with Transitive Dimension and Cyclic Dimension}
Symmetric zero-sum games with two players ($i,j$) is the simplest and extensively studied form of games \cite{15,16}. The joint strategy of the agents is $(\pi_i,\pi_{j})$, and their rewards are $U_i(\pi_i,\pi_{j}) = -U_{j}(\pi_{i},\pi_{j})$. Since the game is symmetric, we do not make distinction between the strategy $\pi$ coming from agent $i$ or $j$. The symmetric property implies that both agents share the same strategy space, and $U_i(\pi^1,\pi^2) = U_j(\pi^2,\pi^1)$. 

Previous studies have shown that the geometric structure of symmetric zero-sum games resembles a spinning top, which consists of the transitive dimension and the cyclic dimension (shown in Figure \ref{fg3}) \cite{5}. The transitive dimension characterizes the absolute strengths between the strategies. A game is transitive if there exists an evaluation function for the strength of the strategy, denoted as $f_v(\pi)$. In the interaction of strategies, the strategy with a higher evaluation function always gets a higher reward:
\begin{equation}
    U_i(\pi^1,\pi^2) = f_v(\pi^1) - f_v(\pi^2) = -U_j(\pi^1,\pi^2).
\end{equation}

The cyclic dimension indicates the presence of mutual restraint among strategies, similar to the dynamics observed in Rock-Paper-Scissor. In a game with only cyclic dimension, for any strategy $\pi^1$ in the strategy space $\Pi$, its expectation of reward when facing other strategies is 0:
\begin{equation}
    \int_{\pi^0 \in \Pi} U_i(\pi^1, \pi^0) \cdot d\pi^0 = 0.
\end{equation}
Real-world games typically exhibit both transitive and cyclic dimensions, leading algorithms based on strategy gradients to often stagnate in cyclic spaces. The primary challenge in symmetric zero-sum games lies in approaching the Nash equilibrium in the transitive dimension.

\subsection{General-Sum Games and Equilibrium Selection Problem}
\label{General-Sum Games and Equilibrium Selection Problem}
Unlike zero-sum games, general-sum games typically feature multiple Nash equilibria with varying rewards \cite{40}. In the context of gradient-based algorithms, distinct initial joint strategies may result in convergence where rewards exhibit notable differences \ref{fg2}. Previous studies suggested the use of specific equilibrium, such as MENE (maximum entropy criterion) \cite{43}, which always exists and is unique in two-player games, or the optimal PSNE (pure strategy Nash equilibrium) \cite{44}. In this paper, our objective is to learn Nash equilibrium with Pareto-optimal rewards \cite{45}.

In the theory of learning in games, agents are typically assigned random initialization and learn strategies with the aim of maximizing their individual rewards. The learning system adopting different algorithms may converge to different equilibrium strategies \cite{20,31}. Previous research has indicated that appropriate guidance for agents can be effective in enhancing their utility of the learning process \cite{8}. In this paper, we will apply the above methods to open-ended frameworks for large-scale general-sum games.

\subsection{Open-Ended Learning Framework}

Prior research has provided many methods for solving Nash equilibrium, including WOLF (Win of Learn Fast) \cite{34} and AWESOME (Adapt When Everybody is Stationary, Otherwise Move to Equilibrium) \cite{33}. The most widely used algorithm is fictitious play \cite{30} due to the simplicity of its execution. 

In fictitious play algorithm with two players, strategies are randomly initialized as $(\pi^0_i,\pi^0_j)$. In iteration $t$, agents select the best response to the average strategy of its opponent:
\begin{equation}
    \pi^{t+1}_i = \operatorname{BR} (\bar{\pi}^t_j), \quad  \bar{\pi}^t_j = \frac{1}{t} \sum^t_{k=1} \pi^k_j.
\end{equation} 
Fictitious play has convergence guarantees in simple structures such as two-player zero-sum games. However, it has the disadvantage that convergence can be very slow in games with large strategy spaces. 

Open-ended algorithm PSRO presents an effective approach to solve Nash equilibrium in games with large-scale strategy spaces. Inspired by the Double Oracle algorithm \cite{21,22}, PSRO establishes a population to represent strategies for each agent. The initial strategy population is generated randomly: $\mathcal{P}_i = (\pi^1_i, \cdots, \pi^t_i)$. In each iteration, the empirical game matrix for agent is calculated as $\mathcal{M}_i = \mathcal{P}_i \times U_i \times \mathcal{P}_{-i}$. By adopting the fictitious play to solve the Nash equilibrium of meta-game, with a payment matrix of $(\mathcal{M}_i,\mathcal{M}_{-i})$, we can derive the meta-equilibrium for agents: $(\theta^*_i,\theta^*_{-i})$. Then, agent $i$ will search for new strategy $\pi^{t+1}_i$, usually the best response to the meta-equilibrium of the opponent $\operatorname{BR}(\theta^*_{-i})$. Pseudo-code for the classic PSRO algorithm is given in the Supplementary Material \ref{Classic PSRO Algorithm}.

Improvements to the PSRO algorithm for solving Nash equilibrium primarily involve the incorporation of a new meta-game solver or the adoption of diverse objectives to guide the generation of new strategies. There are also open-ended algorithms that focus on different equilibrium concepts, such as the $\alpha$-Rank equilibrium ($\alpha$-PSRO) \cite{4} and the correlated equilibrium (JPSRO) \cite{17,46}.

In this paper, we focus on refining the process of searching for new strategies in the PSRO framework. The improved algorithms for PSRO primarily aim at enhancing strategy diversity. As shown in Figure \ref{fg3}, the exploration of the strategy with the best response may fail to approach the Nash equilibrium when facing a strong gradient in the cyclic dimension. To address this, increasing strategy diversity enables the discovery of strategies closer to the Nash equilibrium. There are several methods to measure diversity, including Expected Cardinality (EC) \cite{10}, Behavioral Diversity (BD), and Response Diversity (RD) \cite{2}.

\section{Advantage Policy Space Response Oracle}

\subsection{From Exploitability to Advantage Function}

We first consider symmetric zero-sum games. From the symmetry, we have the following property:
\begin{theorem}
    In symmetric zero-sum games, if the joint strategy $(\pi^1,\pi^2)$ is a Nash equilibrium, we have $(\pi^1,\pi^1)$ and $(\pi^2,\pi^2)$ are both Nash equilibriums.
    \label{th1}
\end{theorem}
From Theorem \ref{th1}, we recognize the importance to establish strategy evaluation metrics for individual agents. However, exploitability can only measure the distance of the joint strategy from the Nash equilibrium. This inspires us to design an extension for the exploitability.

For a strategy $\pi$ in a zero-sum game (which does not require symmetry), its best response $\operatorname{BR}(\pi)$ is usually a set containing many strategies. However, we have the following property:
\begin{theorem}
    For any two-player game, when the strategies of another player is fixed (denoted as $\pi_{j}$), there always exists pure strategy $a_i \in \mathcal{A}$ which satisfies that $a_i \in \operatorname{BR}(\pi_{j})$. Particularly, in zero-sum games, $U_i(\pi_i,\pi_{j})$ is always the same for all $\pi_{j} \in \operatorname{BR}(\pi_i)$.
    \label{th2}
\end{theorem}

Then we have:
\begin{equation}
    \begin{aligned}
        \mathcal{E}(\pi^1,\pi^2) & = \operatorname{max}_{\pi^\prime} U_i(\pi^\prime,\pi^2) - U_i(\pi^1,\pi^2) \\
        & + \operatorname{max}_{\pi^{\prime \prime}} U_j(\pi^1,\pi^{\prime\prime}) - U_j(\pi^1,\pi^2) \\
        & = U_i(\operatorname{BR}(\pi^2) \cap \mathcal{A},\pi^2) + U_j(\pi^1,\operatorname{BR}(\pi^1)\cap \mathcal{A}) \\
        & = - U_j(\operatorname{BR}(\pi^2)\cap \mathcal{A},\pi^2) - U_i(\pi^1,\operatorname{BR}(\pi^1)\cap \mathcal{A})
    \end{aligned}
\end{equation}

If we consider $\operatorname{BR}(\pi^1)$ as a function of $\pi^1$, then the value of $U_i(\pi^1,\operatorname{BR}(\pi^1)\cap \mathcal{A})$ is determined only by $\pi^1$.  
\begin{definition}
In two-player zero-sum games, we define $\mathcal{V}_i(\pi_i) = U_i(\pi_i,\operatorname{BR}(\pi_i))$.
\label{def1}
\end{definition}
From Theorem \ref{th2}, we can see that this definition makes sense because the selection of best response $\operatorname{BR}(\pi_i)$ does not affect the value of $\mathcal{V}$. We find that the definition of advantage has several good properties, which makes it an alternative to the exploitability in zero-sum games.

\begin{theorem}
    In zero-sum games, 
    \begin{itemize}
        \item $\mathcal{E}(\pi_i,\pi_{j}) = - (\mathcal{V}(\pi_i)+\mathcal{V}(\pi_{j}))$.
        \item $\mathcal{V}(\pi)$ is Lipschitz continuous about $\pi$, and $-\mathcal{V}(\pi)$ is a convex function about $\pi$. 
        \item If the game is symmetric, $\forall \pi_i$, $\mathcal{V}(\pi_i) \leq 0$. The joint strategy $(\pi_i,\pi_{i})$ is a Nash equilibrium if and only if $\ \mathcal{V}(\pi_i) = 0$. In games with only transitive dimension, $\mathcal{V}(\pi_i) > \mathcal{V}(\pi_{j})$ implies $U_i(\pi_i,\pi_{j}) > 0$.
    \end{itemize}
    \label{th3}
\end{theorem}
From Theorem \ref{th3}, we can see that the advantage function possesses a unique local and global maximum of 0 in symmetric zero-sum games, which represents the strategy is a Nash equilibrium. This indicates that the process of improving the advantage of strategies results in the convergence to Nash equilibrium. From Theorem \ref{th2}, we can see that the advantage function can be obtained with low computational complexity, because $\operatorname{BR}(\pi_i)$ can be obtained by simply searching in the pure strategy space $\mathcal{A}$.

In the following sections, we will introduce our algorithm for solving the Nash equilibrium with the advantage function as an evaluation metric.

\subsection{A-PSRO for Solving Zero-Sum Games}

As detailed in Section \ref{Symmetric Zero-Sum Games with Transitive Dimension and Cyclic Dimension}, the main challenge in solving the Nash equilibrium in zero-sum games is the existence of transitive and cyclic dimensions, making it difficult for the algorithm to generate new strategies proximate to the Nash equilibrium. In classic PSRO algorithms, new strategies are typically derived through reinforcement learning, with the opponent strategy fixed as the meta-Nash strategy:
\begin{equation}
    \pi^{t+1}_i = \operatorname{BR}(\theta^*_j), \ \text{where} \ (\theta^*_i,\theta^*_j) = \operatorname{NE}(\mathcal{M}_i,\mathcal{M}_j).
\end{equation}

Diversity strategy algorithms offer an improvement over the classic PSRO, with the objective to enhance the diversity of the population during the generation of new strategies. For example, DPP-PSRO \cite{10} includes EC (expected cardinality) as a regularizer to the maximization of the learning player’s own payoff. By doing so, these algorithms increase the probability of discovering novel strategies in the transitive dimension. However, it is important to note that the new strategies generated by diversity algorithms cannot deterministically reduce exploitability.

From Theorem \ref{th3}, we can see that increasing the advantage of strategy will approach the Nash equilibrium. Since $\mathcal{V}(\pi)$ is convex and differentiable almost everywhere, we design the A-PSRO (Advantage Policy Space Response Oracle) to introduce advantage as the objective of strategy learning.

For the population-based strategy update approach in PSRO, we define $\mathcal{V}(\pi \mid \mathcal{P}_i) = U(\pi, \operatorname{BR}^*(\pi)),$ where
\begin{equation}
    \operatorname{BR}^*(\pi) = \operatorname{argmin}_{\pi^\prime \in \mathcal{P}_i} U(\pi,\pi^\prime).
\end{equation}
And we have the following conclusion:
\begin{theorem}
    In symmetric zero-sum games, given the population $\mathcal{P}_{i} = \mathcal{P}_{j} = \{ \pi^1_i,\cdots,\pi^t_i \}$, $\forall \pi^k_i \in \mathcal{P}_i$, we have $\mathcal{V}(\pi^k_i) \leq \mathcal{V}(\theta_i \mid \mathcal{P}_i)$. $\theta_i$ is the equilibrium of the meta-game corresponding to the population $\mathcal{P}_i$.
    \label{th4}
\end{theorem}
It is worth noting that $\mathcal{V}(\pi^k_i) \leq \mathcal{V}(\theta_i)$ does not always hold. However, we have:
\begin{equation}
    \mathcal{V}(\theta_i \mid \mathcal{P}_i) \rightarrow \mathcal{V}(\theta_i), \text{when} \ \operatorname{hull}(\mathcal{P}_i) \rightarrow \Pi,
\end{equation}
where $\Pi$ is the strategy space and $\operatorname{hull}(\mathcal{P}_i)$ is the convex hull of population. From Theorem \ref{th4} we can see that the equilibrium of the meta-game approximately maximizes advantage of the current population. 

We aim to search for new strategy with deterministic increase in the advantage of population. We have the following property of the advantage in population iterations.
\begin{theorem}
    Given the strategy $\theta_i$, if $\mathcal{V}(\theta_i) < 0$, there exists $\Delta \pi_i \in \mathcal{A}$ and $\delta>0$ which satisfies:
    \begin{equation}
        \forall \ 0 < d < \delta, \quad \mathcal{V}\left( \left(1-d\right) \cdot \theta_i + d \cdot \Delta \pi_i \right) >  \mathcal{V}(\theta_i),
    \end{equation}
    \label{th5}
\end{theorem}
Theorem \ref{th5} indicates that if the strategy $\theta_i$ obtained from the current population is not a Nash equilibrium, then we can find a strategy closer to the Nash equilibrium in its neighborhood. We name this process LA (LookAhead). Since we can search the update direction of the strategy in the pure strategy space $\mathcal{A}$, the compotational complexity of doing so is similar to finding the best response. Given step size $d$, the new strategy generated by A-PSRO is:
\begin{equation}
\begin{aligned}
\pi_i^{t+1} & = \left(1-d\right) \cdot \theta_i + d \cdot \Delta \pi_i, \\
\Delta \pi_i & =  \mathop{\text{argmax}}\limits_{\Delta \pi \in \mathcal{A}} \mathcal{V}(\left(1-d\right) \cdot \theta_i + d \cdot \Delta \pi_i).
\end{aligned}
\end{equation}
According to Equation (10), we first calculate the equilibrium of the population $(\theta_i,\theta_j)$. Then we search for the update direction $\Delta \pi_i \in \mathcal{A}$ that maximizes the advantage function with decreasing step size $d$. Our algorithm can deterministically bring the population closer to the Nash equilibrium compared to the randomized search of the diversity strategy approaches. Details of the algorithm and pseudo-code are given in the Supplementary Material \ref{A-PSRO for Solving Zero-Sum Games1}.

\subsection{A-PSRO for Solving Two-Player General-Sum Games}

Definition \ref{def1} can be extended directly to two-player general-sum games. To avoid the existence of best responses with different rewards in general-sum games, we define the simplified game as following:
\begin{definition}
    We call a game simplified game, if for any player $i$ and $\pi_i \in \Pi_i$, there does not exists $a^j_p, a^j_q \in \operatorname{BR}(\pi_i) \cap \mathcal{A}_{j}$ which satisfies that $U_i(\pi_i,a^j_p) \neq U_i(\pi_i,a^j_q)$. 
\label{def2}
\end{definition}

Then we define the advantage function:
\begin{definition}
In two-player general-sum simplified games, we define $\mathcal{V}_i(\pi_i) =  U_i(\pi_i,a_j)$, where $a_j \in \operatorname{BR}(\pi_i) \cap \mathcal{A}_j$.
\label{def3}
\end{definition}

Similar to zero-sum games, the advantage function has the following properties in simplified general-sum games:
\begin{theorem}
    In two-player general-sum simplified games, 
    \begin{itemize}
        \item $\forall i$, $\mathcal{V}_i(\pi_i)$ is Lipschitz continuous. 
        \item We assume that the joint strategy $(\pi_i,\pi_j)$ is a Nash equilibrium. If $\operatorname{BR}(\pi_i)  \cap \mathcal{A}_j = \{a_j^0\}$,  then $\mathcal{V}_i(\pi_i)$ is a local maximum of the advantage function $\mathcal{V}_i$.
        \item If joint strategies $(\pi^1_i,\pi^1_j)$ and $(\pi^2_i,\pi^2_j)$ are both NEs. Then $(\pi^1_i,\pi^1_j)$ Pareto dominates $(\pi^2_i,\pi^2_j)$ if and only if $\mathcal{V}(\pi^1_i) \geq \mathcal{V}(\pi^2_i)$ and $\mathcal{V}(\pi^1_j) \geq \mathcal{V}(\pi^2_j)$.
    \end{itemize}
    \label{th6}
\end{theorem}

From Section \ref{General-Sum Games and Equilibrium Selection Problem} we see that there may be multiple equilibria with different joint rewards in two-player general-sum games. Since the advantage function is connected with the reward of strategies, it can be used to search for equilibrium strategies with higher rewards.

The advantage function is non-convex in general-sum games, which allows the strategy gradient algorithm to converge only to the local maximum. However, when computing the meta-equilibrium $(\theta_i,\theta_j)$ in the population-based PSRO algorithm, we prove that there exists a strategy space that deterministically converges to the global optimum. 
\begin{theorem}
    In two-player general-sum simplified games, the current population is $\mathcal{P}_i=\{\pi^1_i,\cdots,\pi^t_i\}$. $\theta^*_i$ is the global maximum in $\operatorname{hull}(\mathcal{P}_i)$. Then there must exist a non-zero measure set $\mathcal{D}^\prime \subset \operatorname{hull}(\mathcal{P}_i)$, which satisfies that if $\theta^{\prime}_i$ is a local maximum in $\mathcal{D}^\prime$, then $\mathcal{V}(\theta^{\prime}_i) = \mathcal{V}(\theta^{*}_i)$.
    \label{th7}
\end{theorem}
Theorems \ref{th6} and \ref{th7} state that there exists an non-zero measure set near the optimal equilibrium in general-sum games, in which the population of strategies converges in the direction of that equilibrium. Strategies near equilibria with optimal rewards have a higher advantage due to the Lipschitz continuity of the advantage function. In A-PSRO, we adopt the following approach in strategy exploring to increase the probability of converging to a better equilibrium. 
\begin{equation}
\begin{aligned}
    & \pi^{t+1}_i  = (1-d) \cdot \theta_i + d \cdot \Delta \pi_i, \\
    & \theta_i  = \mathop{\text{argmax}}\limits_{(\theta^\prime_i,\theta^\prime_j) \in \Theta} \ \mathcal{V}_i(\theta^\prime_i), \ \Theta = \bigcup\limits_{\pi_i \in \operatorname{hull}(\mathcal{P}_i)} \mathcal{O}(\mathcal{P}_i, \mathcal{P}_j \mid \pi_i).
\end{aligned}
\end{equation}
Here, $d$ is the fixed step size, the calculation of $\Delta \pi_i$ is the same as Equation (10). $\mathcal{O}(\mathcal{P}_i, \mathcal{P}_j \mid \pi_i)$ represents the equilibrium obtained through a fictitious play oracle with $\pi_i$ as initial strategy. The computation of $\theta_i$ with best advantage in Equation (11) is difficult, and we have found in experiments that a small amount of oracle is enough to obtain a valid approximation. Details of the algorithm and the pseudo-code are given in the Supplementary Material \ref{A-PSRO for Solving Two-Player General-Sum Games1}.

\begin{figure*}[t]
\centering
\subfigure[AlphaStar] {\label{fg1_1}
\includegraphics[width=0.49\columnwidth]{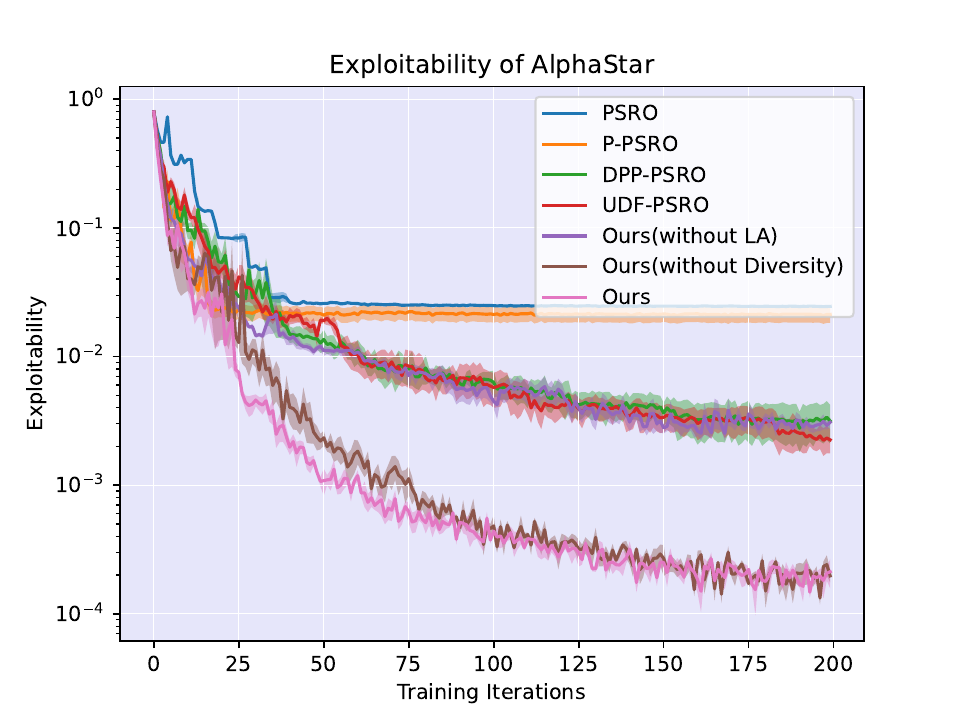}
}
\subfigure[Transitive game] {\label{fg1_2}
\includegraphics[width=0.49\columnwidth]{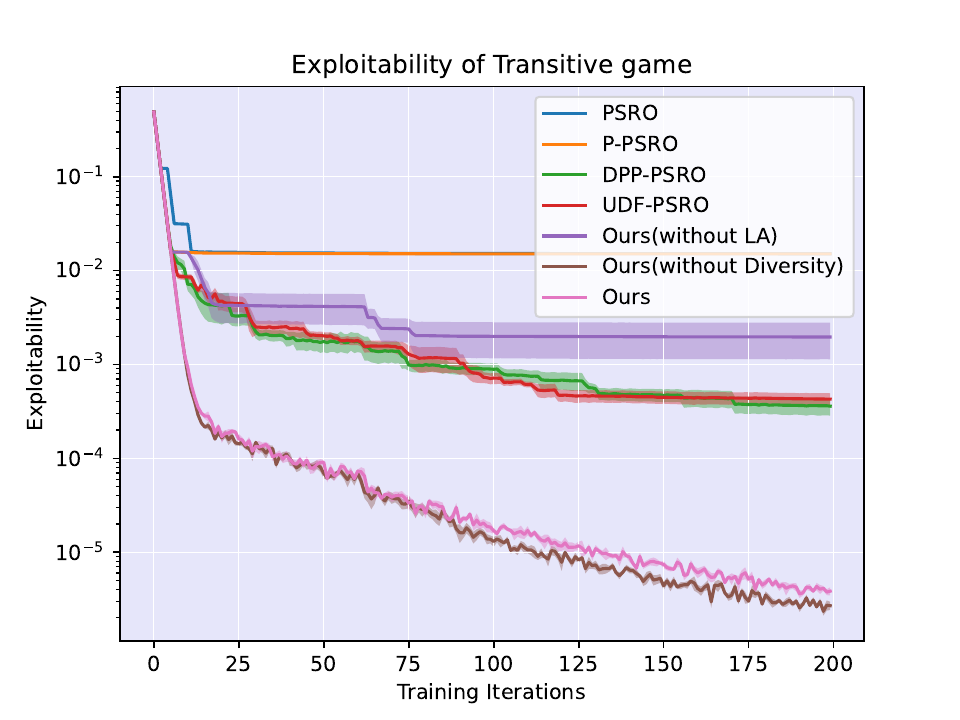}
}
\subfigure[Elo game $+$ noise=1.0] {\label{fg1_3}
\includegraphics[width=0.49\columnwidth]{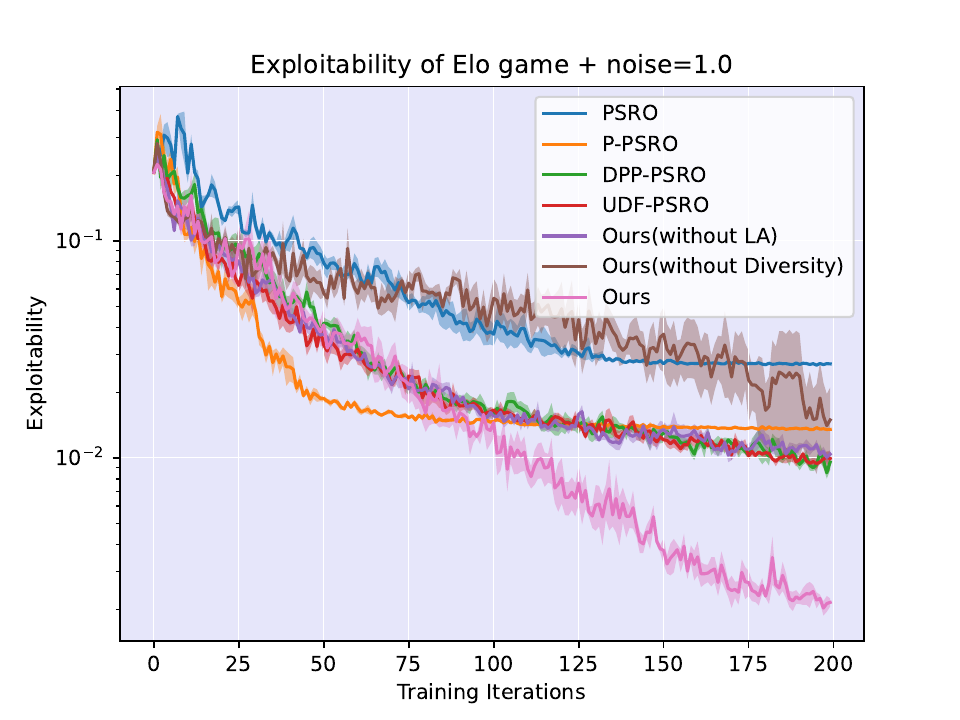}
}
\subfigure[Simplified Go game] {\label{fg1_4}
\includegraphics[width=0.49\columnwidth]{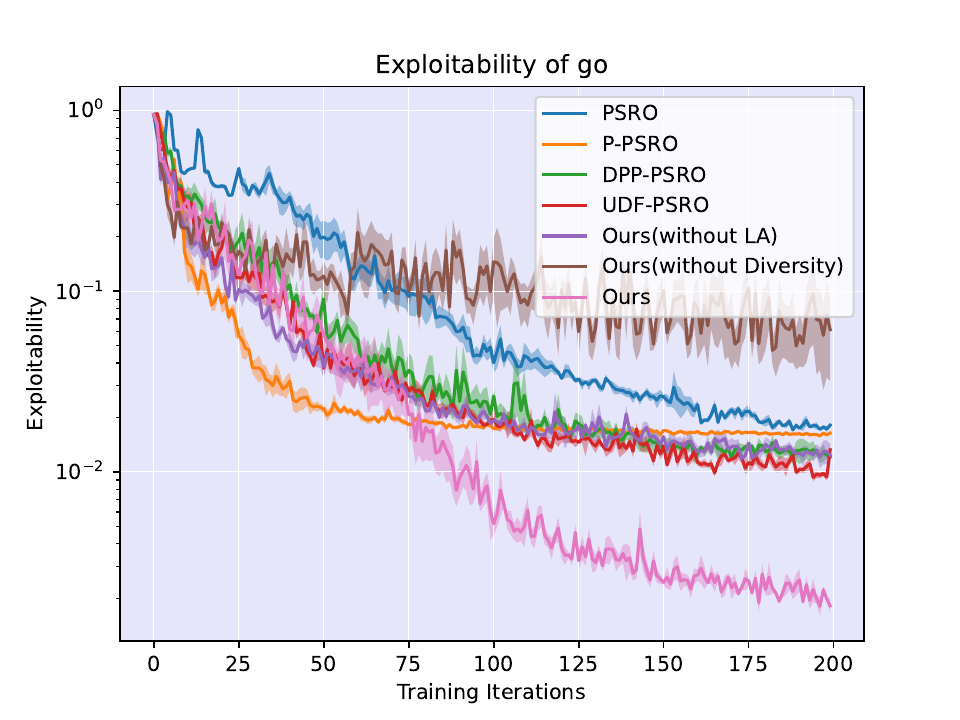}
}
\vspace{-0.5cm}
\caption{The exploitability of the joint strategy learned by agents in various zero-sum games is depicted. The reduction in exploitability through population iterations can serve as an indicator of the effectiveness in approximating the Nash equilibrium.}
\label{fig3}
\vspace{-0.5cm}
\end{figure*}

\begin{figure*}[t]
\centering
\subfigure[AlphaStar] {\label{fg2_1}
\includegraphics[height=0.38\columnwidth,width=0.49\columnwidth]{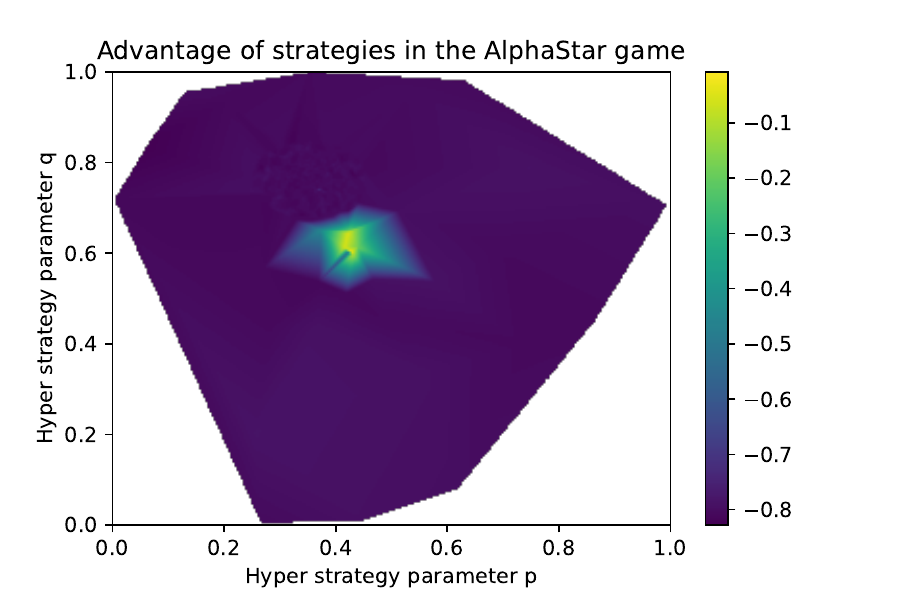}
}
\subfigure[Transitive game] {\label{fg2_2}
\includegraphics[height=0.38\columnwidth,width=0.49\columnwidth]{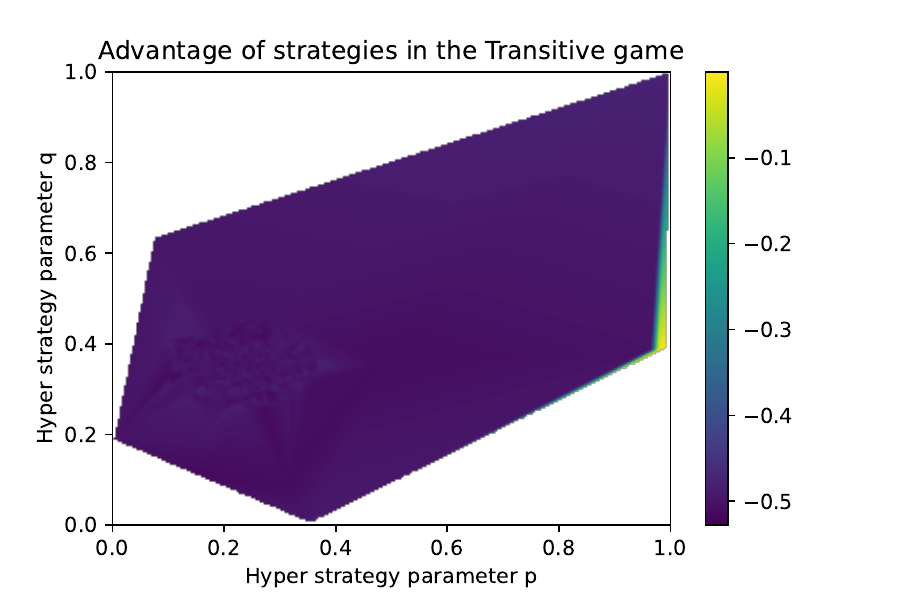}
}
\subfigure[Elo game $+$ noise=1.0] {\label{fg2_3}
\includegraphics[height=0.38\columnwidth,width=0.49\columnwidth]{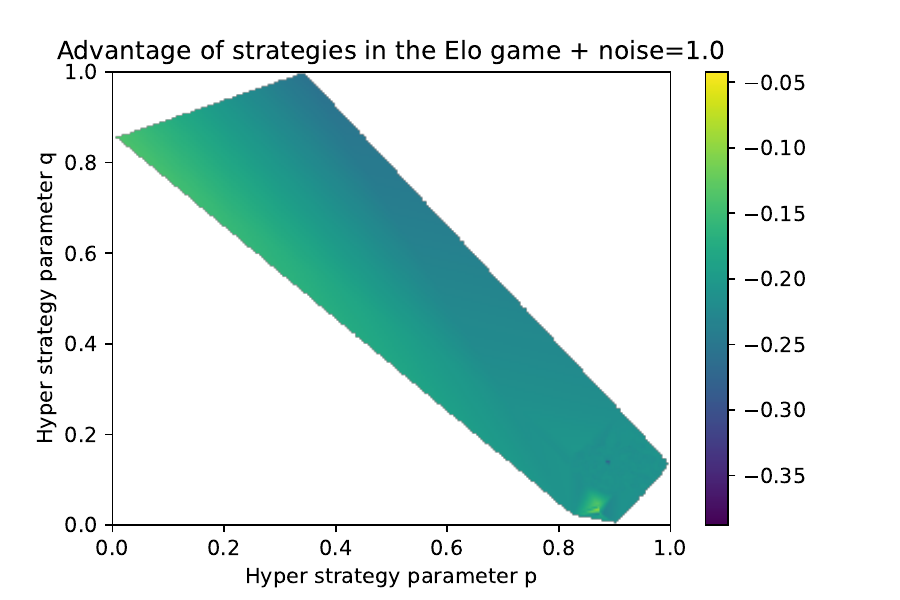}
}
\subfigure[Simplified Go game] {\label{fg2_4}
\includegraphics[height=0.38\columnwidth,width=0.49\columnwidth]{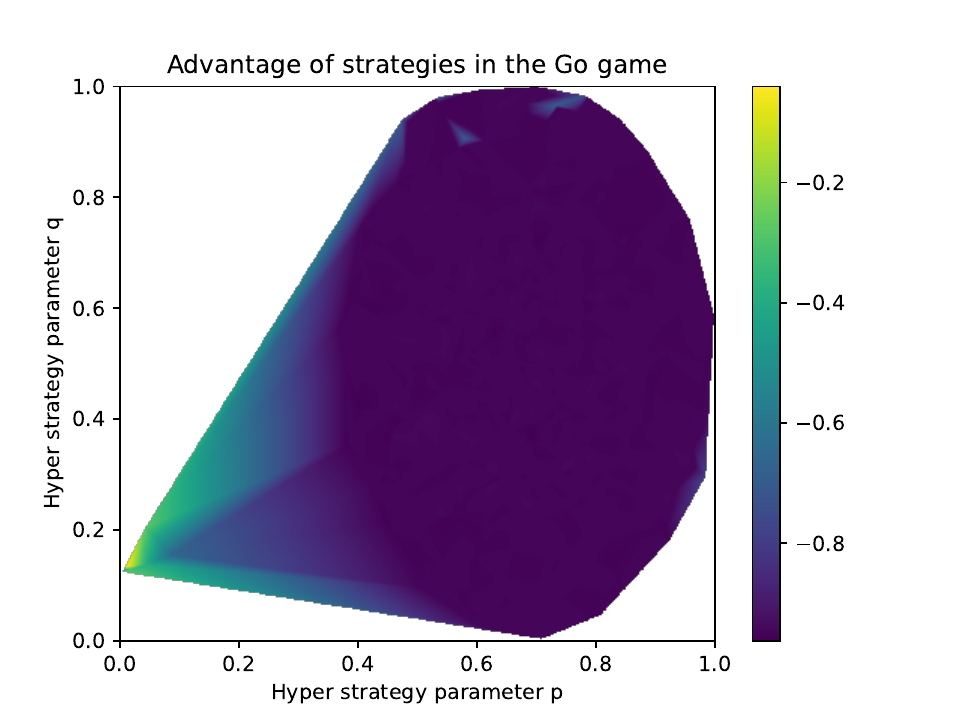}
}
\vspace{-0.4cm}
\caption{The advantage of strategies in different games. Strategies with lighter colors have a higher advantage. In symmetric zero-sum games, the Nash equilibrium strategy has the highest advantage 0.}
\label{fig4}
\vspace{-0.5cm}
\end{figure*}

\begin{figure}[t]
\centering
\subfigure[AlphaStar] {\label{fg3_1}
\includegraphics[width=0.47\columnwidth]{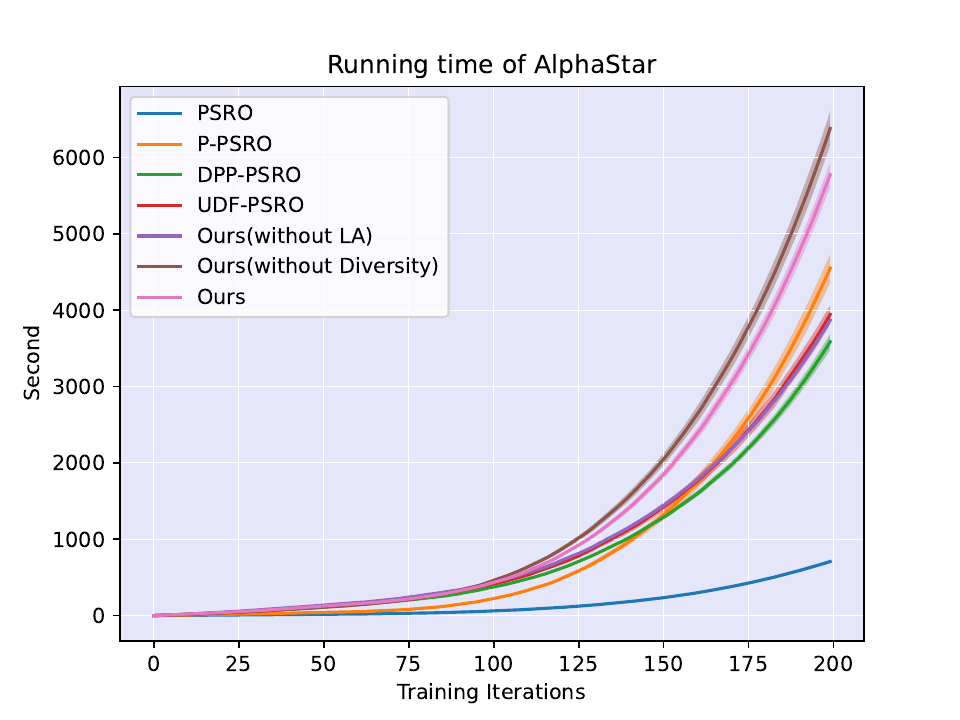}
}
\subfigure[Simplified Go game] {\label{fg3_2}
\includegraphics[width=0.47\columnwidth]{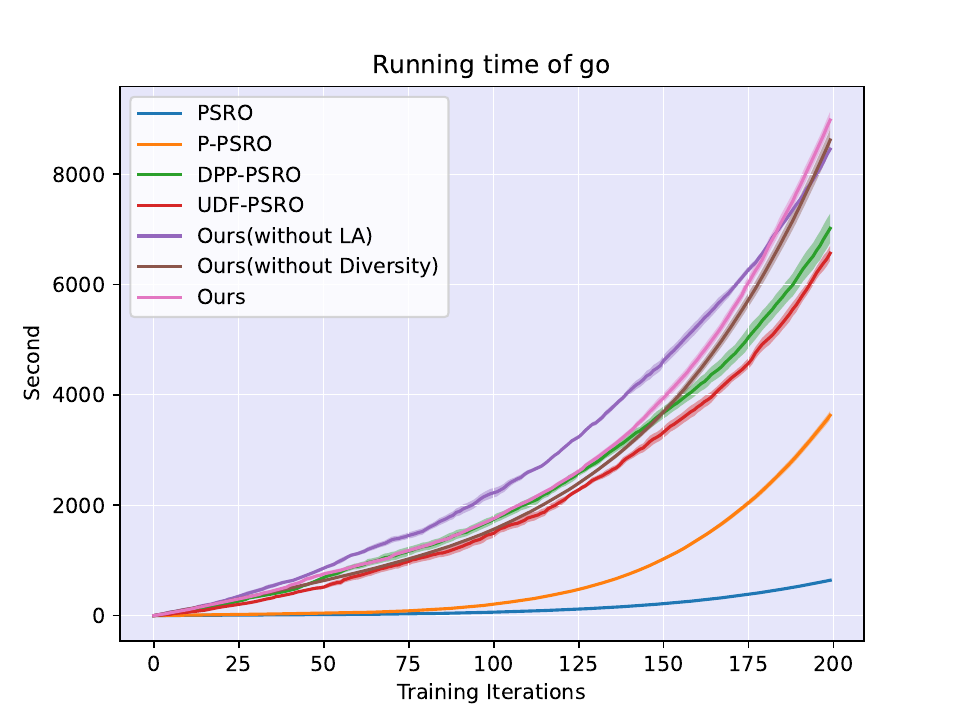}
}
\vspace{-0.4cm}
\caption{The running time of different algorithms.}
\label{fig5}
\vspace{-0.5cm}
\end{figure}

\subsection{A-PSRO for Solving Multi-Player Games}
In the multi-player general-sum games, the definiton of the advantage function cannot be obtained from the best response. We define the advantage function as:
\begin{equation}
    \mathcal{V}(\pi_i) = U_i(\pi_i,\mu(\pi_i)), \ \mu(\pi_i) = \mathcal{O}(\Pi_{-i} \mid \pi_i). 
\end{equation}
In this definition, $\mu(\pi_i)$ is joint strategy without player $i$ as the equilibrium of the subgame when the strategy of player $i$ is $\pi_i$. The computation of $\mu(\pi_i)$ comes from the equilibrium oracle $\mathcal{O}$. In symmetric zero-sum games, the equilibrium of the subgame is the same for the best response $\operatorname{BR}(\pi_i)$. 

This definition is similar to the multi-follower stackelberg game \cite{41}. Different equilibrium oracle can result in different advantage functions. In multi-follower stackelberg games, commonly used oracles is the pessimistic equilibrium \cite{42}. Taking the pessimistic equilibrium oracle as example:
\begin{equation}
    \begin{aligned}
    & \mu(\pi_i) = \mathop{\text{argmin}}\limits_{\pi_{-i}} U_i(\pi_i, \pi_{-i}) \\
    & \text{s.t.} \  \pi_j \in \mathop{\text{argmax}}\limits_{\pi_j} U_j(\pi_j \mid \pi_{-j}), \ \forall j \in \mathcal{N} \setminus \{i\}. 
    \end{aligned}    
\end{equation}
We give a simple approximation of the pessimistic equilibrium oracle. We view the other agents as a single agent $\{-i\}$ with action space $\mathcal{A}_{-i}$, and choose the worst-case joint action in their joint action space:
\begin{equation}
    \mu(\pi_i) = \mathop{\text{argmin}}\limits_{\pi_{-i} \in \mathcal{A}_{-i}} U_i(\pi_i, \pi_{-i})
\end{equation}
We present the computation of the advantage function in multi-player game in Supplementary Material \ref{A-PSRO for Solving Multi-Player Games1}. In multi-player games, we establish a similar framework to Equations (10) and (11) for A-PSRO. The rest of the population iteration framework is consistent with the two-player game. In our experiments for multi-player games, we find that A-PSRO demonstrates the same effectiveness.

\begin{figure*}[t]
\centering
\subfigure[Advanced-Staghunt] {\label{fg4_1}
\includegraphics[width=0.65\columnwidth]{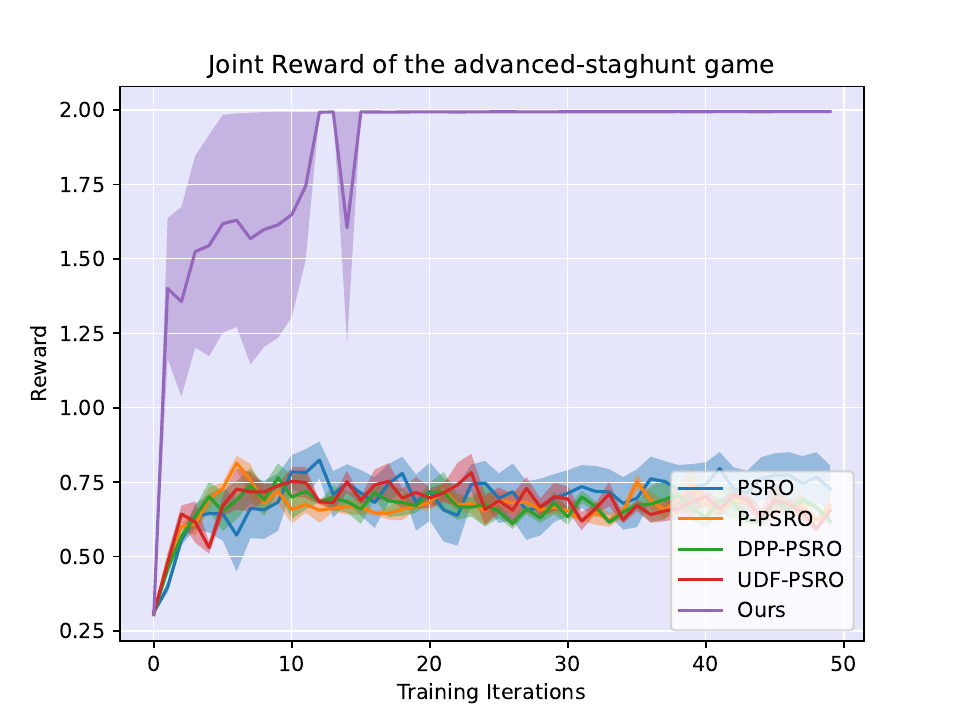}
}
\subfigure[Advanced-RSP] {\label{fg4_2}
\includegraphics[width=0.65\columnwidth]{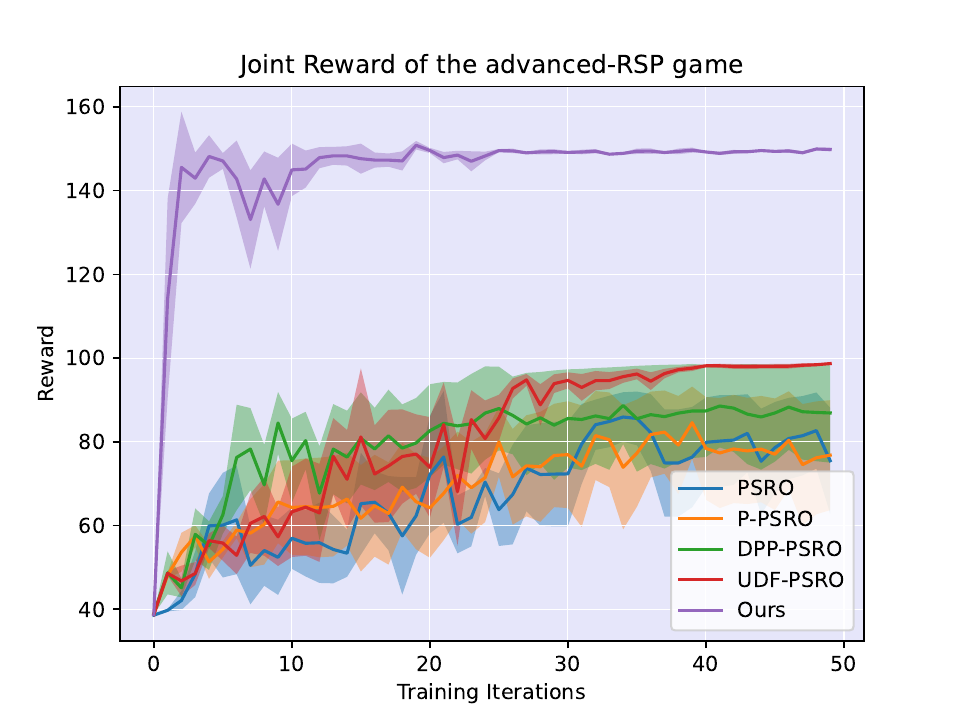}
}
\subfigure[Randomly Generated Games] {\label{fg4_3}h
\includegraphics[width=0.65\columnwidth]{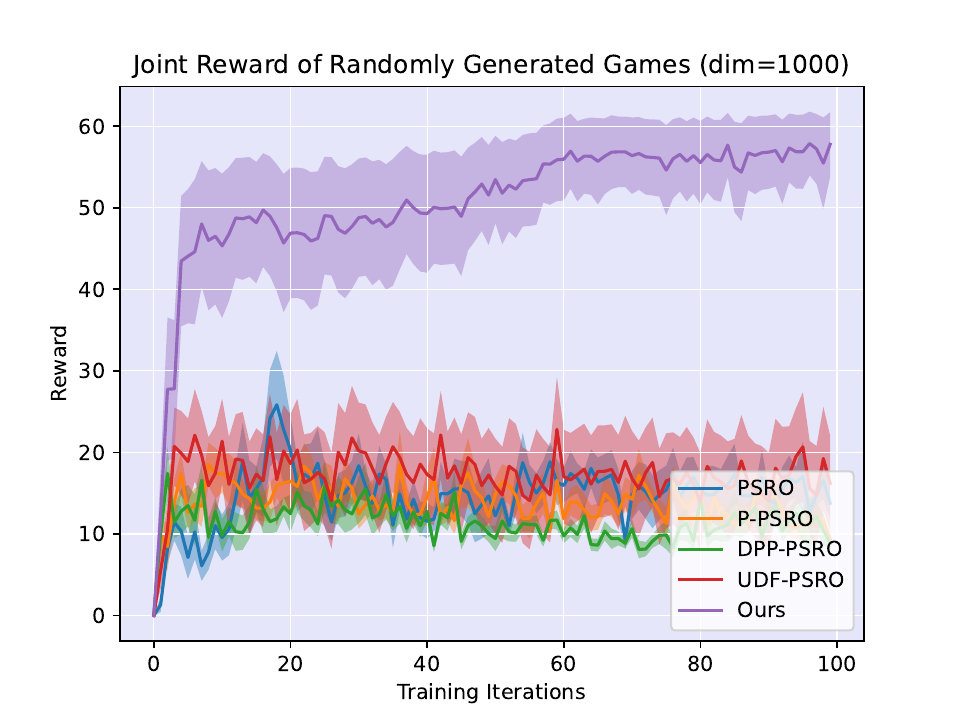}
}
\vspace{-0.2cm}
\caption{The joint reward of the agent system in general-sum games. The Staghunt game and the RSP game are repeated 10 times and averaged for plotting. Randomly generated games contain 100 games with the same reward distribution.}
\label{fig6}
\vspace{-0.4cm}
\end{figure*}

\begin{figure*}[t]
\centering
\subfigure[Advanced-Staghunt] {\label{fg5_1}
\includegraphics[width=0.66\columnwidth]{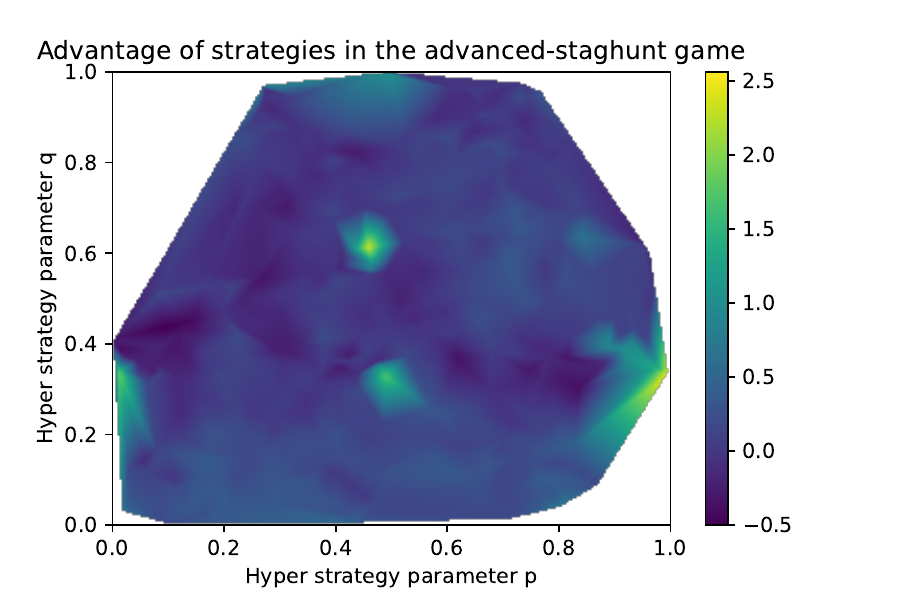}
}
\subfigure[Advanced-RSP] {\label{fg5_2}
\includegraphics[width=0.66\columnwidth]{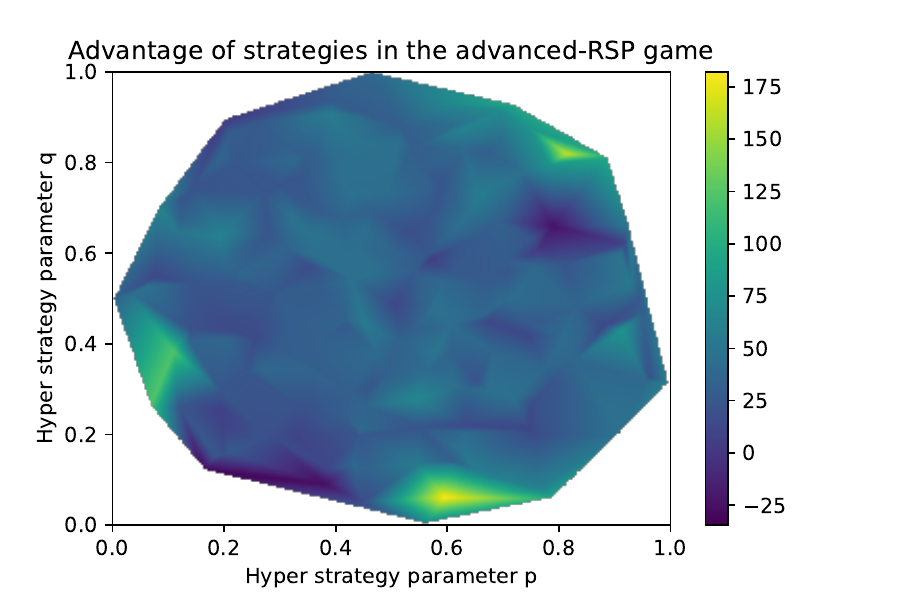}
}
\subfigure[Randomly Generated Games] {\label{fg5_3}
\includegraphics[width=0.66\columnwidth]{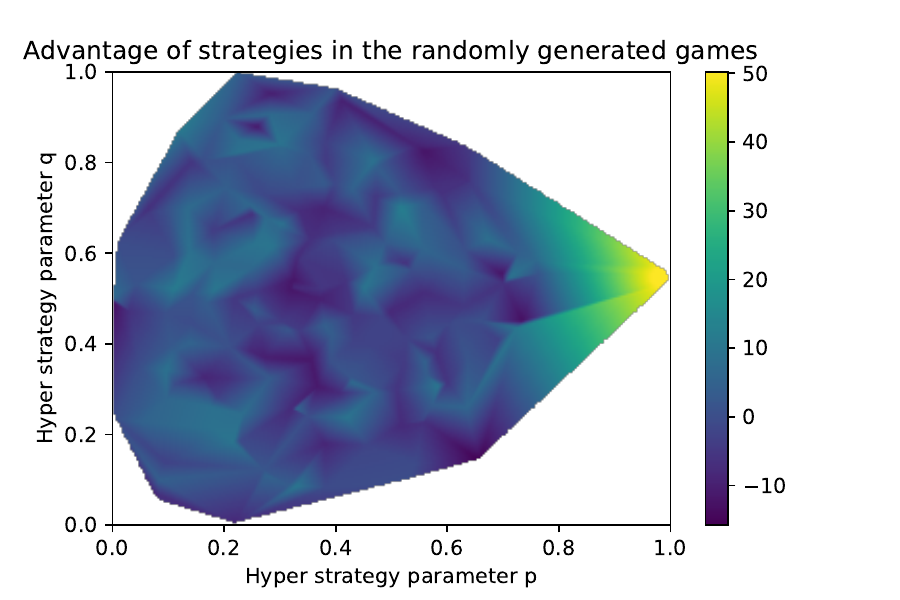}
}
\vspace{-0.2cm}
\caption{The advantage distribution of strategies in different two-player general-sum games.}
\label{fig7}
\vspace{-0.4cm}
\end{figure*}

\section{Experiment Results and Discussion}

We evaluate the performance of A-PSRO in multiple game environments. We select the state-of-the-art game solvers as baselines, including PSRO \cite{1}, Pipeline-PSRO (P-PSRO) \cite{9}, DPP-PSRO \cite{10} and UDF-PSRO \cite{2}.

\subsection{Experiments in Symmetric Zero-Sum Games}
In symmetric zero-sum games, we test A-PSRO with both LookAhead and diversity modules on complex real-world games and report the exploitability of different algorithms. The environment we chose for testing is the normal-form games generalized used in the previous PSRO algorithms \cite{5,3}. These games are derived from real-world large-scale games abstracted into matrix form, which are described in detail in the Supplementary Metarial \ref{A-PSRO for Solving Zero-Sum Games2}.  

In Figure \ref{fig3}, we show the results in four typical zero-sum games: AlphaStar (the widely used real-world game with a complex strategy space), Transitive game (game with almost only transitive dimension), Elo game with noise (noise is randomly added to the payment matrix), and Simplified Go game (traditional board game with restricted board size). The results of more games are presented in the Supplementary Material, Figure \ref{supply1}.

From Figure \ref{fig3}, we can see that our method achieves a notable reduction in exploitability across all game environments, sometimes by several orders of magnitude. A-PSRO without diversity outperforms A-PSRO in the Transitive game. The reason is that the diversity module is designed to navigate the constraints of non-transitive structure, and its impact is limited in games with strong transitive dimensions. The effectiveness of A-PSRO without the LookAhead module has a significant decline in all the games, which indicates that our LookAhead module greatly contributes to approximating Nash equilibria in both cyclic and transitive structural games.

In Figure \ref{fig4}, we show the distribution of advantages in different games. We use the non-linear dimensionality reduction method t-SNE (t-Distributed Stochastic Neighbor Embedding) \cite{39} to map the strategies into the unit matrix and maintain adjacency between strategies. 

From Figure \ref{fig4}, it is evident that there exists clear convexity of the advantage function in zero-sum games. In games with a strong transitive dimension (AlphaStar, Transitive game), the advantage function exhibits rapid changes around the Nash equilibrium. Conversely, when the advantage function changes slowly around the Nash equilibrium, indicative of a cyclic structure, the diversity module becomes crucial for the learning process of the Nash equilibrium. This is consistent with the experimental results in Figure \ref{fg1_3}, \ref{fg1_4}. 

As A-PSRO solely requires traversing the pure strategy space when computing advantages, the algorithm can be swiftly implemented using matrix operations. Figure \ref{fig5} illustrates the comparison of running times between different algorithms. It is evident that A-PSRO does not substantially increase computational complexity.

\begin{figure*}[t]
\centering
\subfigure[Exploitability of Zero-Sum Multiagent Games] {\label{fg6_1}
\includegraphics[width=0.80\columnwidth]{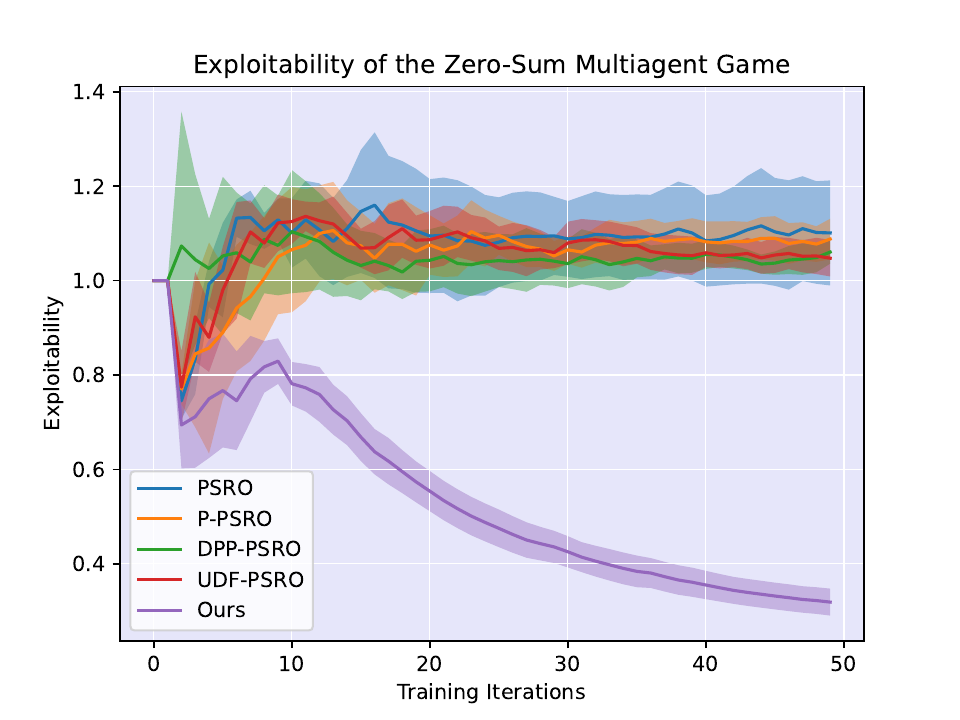}
}
\subfigure[Joint Reward of General-Sum Multiagent Games] {\label{fg6_2}
\includegraphics[width=0.88\columnwidth]{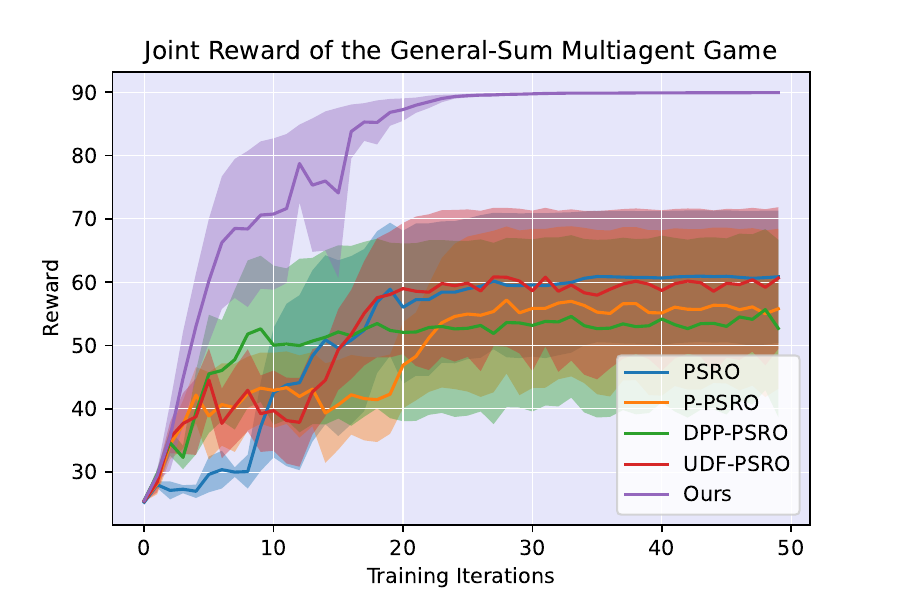}
}
\vspace{-0.2cm}
\caption{The exploitability of zero-sum multiagent games and joint reward of general-sum multiagent games. Each algorithm is tested in 4 randomly generated identically distributed game environments, and the averages are plotted.}
\label{fig8}
\vspace{-0.4cm}
\end{figure*}

\subsection{Experiments in Two-player General-Sum Games}

In two-player general-sum games, we evaluate our method using both traditional games and randomly generated games. Given the absence of generalized test sets for large-scale general-sum games in prior studies, we extend the classic Staghunt game \cite{40} with an equilibrium selection problem and RSP for testing purposes. The detailed game structure is given in the Supplementary Material \ref{A-PSRO for Solving Two-Player General-Sum Games2}.

In Figures \ref{fg4_1} and \ref{fg4_2}, we present the training results of the algorithms in the aforementioned games. To assess the ability of algorithms to learn strategies with higher rewards, we employ the joint reward in the converged state of the algorithm as an evaluation metric. Figure \ref{fig6} demonstrates that A-PSRO consistently learns the optimal equilibrium strategy, whereas other algorithms acquire different equilibria and often stagnate in suboptimal equilibria.

We further conduct experiments in randomly generated games with the same reward distribution, and the results are depicted in Figure \ref{fg4_3}. In our experiments, A-PSRO consistently attains the highest reward, underscoring its suitability for large-scale general-sum games.

Similarly to Section 4.1, we depict the distribution of advantage in the aforementioned games in Figure \ref{fig7}. It is evident from the figure that the advantage of the general-sum game is non-convex. Algorithms based on the strategy gradient are likely to converge to different equilibrium strategies from various initial points.

\subsection{Experiments in Multi-player Games}

For multi-player games, we test our method in both zero-sum and general-sum games. We randomly generated a series of identically distributed games as test environments. The specific environment parameters are given in the Supplementary Material \ref{A-PSRO for Solving Multi-Player Games2}.

In Figure \ref{fg6_1}, the distance of the learned strategies of different algorithms from the Nash equilibrium in multi-player zero-sum games is illustrated. From the figure, it is apparent that existing algorithms do not efficiently incorporate more effective strategies in the process of population expansion. Due to the high solution complexity of multi-player games, this results in the phenomenon that the exploitability of the algorithm increases with population expansion. In contrast, A-PSRO effectively explores strategies with higher advantage and approaches the Nash equilibrium during the training iterations.

Figure \ref{fg6_2} presents the joint rewards of the algorithms during the training process of multi-player general-sum games. The experimental results align with Figure \ref{fg4_3}, indicating that A-PSRO consistently learns the equilibrium strategy with the highest joint reward. In contrast, other PSRO algorithms occasionally stagnate in suboptimal equilibria. These experimental results illustrate that A-PSRO can efficiently learn equilibrium strategies in multi-player games as well.

\section{Conclusion}

In this paper, we introduce A-PSRO, a unified open-ended framework for learning strategies in normal-form games. We propose the advantage function as an evaluation metric for the strategy learning of agents. The advantage function exhibits favorable properties, such as convexity and Lipschitz continuity. Leveraging the advantage function, our algorithm, A-PSRO, effectively enhances the objective of strategy exploration during population expansion. In zero-sum games characterized by complex transitive and cyclic structures, A-PSRO can deterministically approach Nash equilibrium strategies during iterations, significantly reducing the exploitability of learned strategies. Moreover, in general-sum games with multiple equilibria, A-PSRO maximizes rewards during the learning of Nash equilibria, thereby averting stagnation in suboptimal solutions. Experimental results demonstrate the robust generalization capabilities of A-PSRO as an open-ended framework in large-scale environments, highlighting its potential to advance equilibrium theory in multiagent systems.

\clearpage

\nocite{langley00}

\bibliography{example_paper}
\bibliographystyle{icml2024}

\newpage
\appendix
\onecolumn
\section{Omitted Proofs}

\subsection{Proof of Theorem 3.1}
\begin{theorem}
    In symmetric zero-sum games, if the joint strategy $(\pi^1,\pi^2)$ is a Nash equilibrium, we have $(\pi^1,\pi^1)$ and $(\pi^2,\pi^2)$ are both Nash equilibriums.
\end{theorem}

\begin{proof}
From the definition, the joint strategy $(\pi^1,\pi^2)$ is a Nash equilibrium implies that the exploitability $\mathcal{E}(\pi^1,\pi^2)=0$. Then we will have:
\begin{equation}
    \mathcal{E}(\pi^1, \pi^2)   = \operatorname{max}_{\pi^{\prime}_i} [U_i(\pi^{\prime}_i, \pi^2) - U_i(\pi^1, \pi^2)] 
     + \operatorname{max}_{\pi^{\prime}_j} [U_{j}(\pi^1,\pi^{\prime}_j) - U_{j}(\pi^1,\pi^{2})]  
     = 0.
\end{equation}
This indicates that:
\begin{equation}
    \operatorname{max}_{\pi^{\prime}_i} [U_i(\pi^{\prime}_i, \pi^2) - U_i(\pi^1, \pi^2)] = \operatorname{max}_{\pi^{\prime}_j} [U_{j}(\pi^1,\pi^{\prime}_j) - U_{j}(\pi^1,\pi^{2})]  
     = 0.
\end{equation}

Then we prove that $U_i(\pi^1,\pi^2) = U_j(\pi^1,\pi^2) = 0$. If the reward of agents are not all 0, since the game is zero-sum, we assume that:
\begin{equation}
    U_i(\pi^1, \pi^{2}) > U_{j}(\pi^1, \pi^{2}).
\end{equation}
Since the game is symmetric, we will have:
\begin{equation}
    U_i(\pi^1, \pi^1) = U_{j}(\pi^1, \pi^1) = 0.
\end{equation}    
This indicates that:
\begin{equation}
    \operatorname{max}_{\pi^{\prime}_j} [U_{j}(\pi^1,\pi^{\prime}_j) - U_{j}(\pi^1,\pi^{2})] \neq 0,
\end{equation}    
which leads to contradiction. Therefore, we prove that the rewards of both agents are 0.

From the equations above we have that $U_j(\pi^1,\pi^1) = U_j(\pi^1,\pi^2) = 0$. Since $\pi^2$ is a best response to $\pi^1$, we can see that $\pi^1$ is also a best response to $\pi^1$. This indicates that $(\pi^1,\pi^1)$ and $(\pi^2,\pi^2)$ are both Nash equilibriums.
\end{proof}


\subsection{Proof of Theorem 3.2}

\begin{theorem}
    For any two-player game, when the strategies of another player is fixed (denoted as $\pi_{j}$), there always exists pure strategy $a_i \in \mathcal{A}$ which satisfies that $a_i \in \operatorname{BR}(\pi_{j})$. Particularly, in zero-sum games, $U_i(\pi_i,\pi_{j})$ is always the same for all $\pi_{j} \in \operatorname{BR}(\pi_i)$.
\end{theorem}

\begin{proof}
We assume that there exists strategy $\pi^*_i=(p_1,\cdots,p_m)$ which is the best response of $\pi_j$. Then we have:
\begin{equation}
    U_i(\pi^*_i,\pi_j) = p_1 U_i(a_1,\pi_j) + \cdots + p_m U_i(a_m,\pi_j) \leq \operatorname{max}_{l \in \{1,\cdots,m\}} U_i(a_l,\pi_j),
\end{equation}
which implies that $a_l \in \mathcal{A}$ is a best response to $\pi_j$.

In zero-sum games, we assume that the strategy of the player $i$ is set as $\pi_i$. Then we have:
\begin{equation}
    \operatorname{BR} (\pi_i) = \operatorname{argmax}_{\pi_{j}} U_{j} (\pi_{i},\pi_j).
\end{equation}
We assume that for all $\pi_{j} \in \operatorname{BR}(\pi_i)$, the reward of player $j$ is $U_{j}(\pi_i,\pi_j) = U^0$. If the game is zero-sum, the reward of player $i$ is $U_i(\pi_i,\pi_j) = -U^0$. This implies that in zero-sum games, $U_i(\pi_i,\pi_{j})$ is always the same for all $\pi_{j} \in \operatorname{BR}(\pi_i)$.
\end{proof}


\subsection{Proof of Theorem 3.4}

\begin{theorem}
    In zero-sum games, 
    \begin{itemize}
        \item $\mathcal{E}(\pi_i,\pi_{j}) = - (\mathcal{V}(\pi_i)+\mathcal{V}(\pi_{j}))$.
        \item $\mathcal{V}(\pi)$ is Lipschitz continuous about $\pi$, and $-\mathcal{V}(\pi)$ is a convex function about $\pi$. 
        \item If the game is symmetric, $\forall \pi$, $\mathcal{V}(\pi) \leq 0$. The joint strategy $(\pi^1,\pi^{1})$ is a Nash equilibrium if and only if $\ \mathcal{V}(\pi^1) = 0$. In games with only transitive dimension, $\mathcal{V}(\pi^1) > \mathcal{V}(\pi^{2})$ implies $U_i(\pi^1,\pi^2) > 0$.
    \end{itemize}
\end{theorem}

\begin{proof}
\begin{itemize}
\item From the definition, we can easily find that $\mathcal{E}(\pi_i,\pi_{j}) = - (\mathcal{V}(\pi_i)+\mathcal{V}(\pi_{j}))$. We define the domain of strategies $D = \{(p_1,p_2,\cdots,p_m) \}$ which satisfies that $p_k \geq 0$, $\sum^m_{k=1} p_k = 1$, and $m$ is the dimension of the action space $\mathcal{A}=\{a_1,\cdots,a_m\}$. 

\item In order to prove that $-\mathcal{V}(\pi)$ is convex function about $\pi$, we need to prove that for $\pi^1$,$\pi^2 \ \in D$, and $\theta \in (0,1)$, we have $-\mathcal{V}[(1-\theta)\pi^1 + \theta \pi^2] \leq -(1-\theta)\mathcal{V}(\pi^1) - \theta \mathcal{V}(\pi^2)$.

We assume that $\pi^1 = (p^1_1,\cdots,p^1_m)$ and $\pi^2 = (p^2_1,\cdots,p^2_m)$. Then we have:
\begin{equation}
\begin{aligned}
    \mathcal{V}[(1-\theta)\pi^1+\theta \pi^2] & = U[(1-\theta)\pi^1+\theta \pi^2, a_0], \quad a_0 \in \mathcal{A} \cap \operatorname{BR}[(1-\theta)\pi^1+\theta \pi^2] \\
    & = (1-\theta)U(\pi^1,a_0) + \theta U(\pi^2,a_0) \\
    & \geq (1-\theta)U(\pi^1,a^1_0) + \theta U(\pi^2,a^2_0), \quad a^t_0 \in \mathcal{A} \cap \operatorname{BR}(\pi^t), \ t \in \{1,2\}  \\
    & = (1-\theta) \mathcal{V}(\pi^1) + \theta \mathcal{V}(\pi^2).
\end{aligned}
\end{equation}
This implies that the inverse function of advantage function $-\mathcal{V}(\pi)$ is convex about $\pi$. 

Then we prove that $\mathcal{V}(\pi)$ is Lipschitz continuous about $\pi$. We assume that $\pi = (p_1, \cdots, p_m)$, $\pi^\prime = (p_1 + \Delta p_1, \cdots, p_m + \Delta p_m)$ which satisfies that $\sum^m_{k=1} \Delta p_k = 0$, and $a^\prime_0 \in \mathcal{A} \cap \operatorname{BR}(\pi + \Delta \pi), \ a_0 \in \mathcal{A} \cap \operatorname{BR}(\pi)$.

\begin{equation}
\begin{aligned}
   & \mathcal{V}(p_1 + \Delta p_1, \cdots, p_m + \Delta p_m) - \mathcal{V}(p_1, \cdots, p_m) \\
   = & U(\pi + \Delta \pi, a^\prime_0) - U(\pi, a_0) \\
   = & (p_1 + \Delta p_1) U(a_1, a^\prime_0) + \cdots + (p_m + \Delta p_m) U(a_m, a^\prime_0) - p_1 U(a_1, a_0) - \cdots - p_m U(a_n, a_0) \\
   = & A_1 \Delta p_1 + \cdots + A_m \Delta p_m + [U(\pi,a^\prime_0)-U(\pi,a_0)]
\end{aligned}
\end{equation}
where $A_1,\cdots,A_m$ are constants.

Then we will prove that there exists $M$ such that $\forall \delta > 0$, the following conclusion holds:

\begin{equation}
    \lvert U(\pi,a^\prime_0)-U(\pi,a_0) \rvert <  \delta M, \quad \text{when} \ \operatorname{max} \lvert \Delta p_k \rvert \leq \delta.
\end{equation}

First we consider that $a^\prime_0 \in \mathcal{A} \cap \operatorname{BR}(\pi)$, this indicates that $\lvert U(\pi, a^\prime_0) - U(\pi, a_0) \rvert = 0$.

Then we consider that $a^\prime_0 \notin \mathcal{A} \cap \operatorname{BR}(\pi)$, which means that:

\begin{equation}
    U(\pi,a_0) > U(\pi,a^\prime_0) 
\end{equation}

We prove that $U(\pi,a)$ is Lipschitz continuous with respect to $\pi$. $\forall \pi, \forall a$, we have:

\begin{equation}
\begin{aligned}
    U(\pi+\Delta \pi,a) - U(\pi,a) & = \sum^m_{k=1} (p_k+\Delta p_k) U (a_k,a) - \sum^m_{k=1} p_k U(a_k,a) \\
    & = \Delta p_1 U(a_1, a) + \cdots + \Delta p_m U(a_m,a) \\
    & \leq \delta M \quad \text{where} \ \delta = \operatorname{max} \lvert \Delta p_k \rvert \ \text{and} \ M = \mathop{\text{max}}\limits_{a^* \in \mathcal{A}} \ \lvert \sum^m_{i=1} U(a_i,a^*) \rvert.
\end{aligned}
\end{equation}
Since we are considering games with finite dimension, M has consistent upper bounds. Then we have:

\begin{equation}
    U(\pi, a^\prime_0) + \delta M \geq U(\pi^\prime,a^\prime_0) \geq U(\pi^\prime,a_0) \geq U(\pi,a_0) - \delta M.
\end{equation}

This indicates that $0 \leq U(\pi,a_0) - U(\pi, a^\prime_0) \leq 2 \delta M$, which means that $\mathcal{V}(\pi)$ is Lipschitz continuous about $\pi$.

\item From Theorem 3.1 we can easily find that if the game is symmetric, $\mathcal{V}(\pi) \leq 0$. The joint strategy $(\pi^1,\pi^1)$ is a Nash equilibrium if and only if $\ \mathcal{V}(\pi^1) = 0$.

In games with only transitive dimension, the best response is the same $\pi^0$ for all strategies $\pi$. Then we have:
\begin{equation}
    \mathcal{E}(\pi^1,\pi^0) = -\mathcal{V} (\pi^1) < -\mathcal{V} (\pi^2) = \mathcal{E}(\pi^2,\pi^0)
\end{equation}
So $\pi^1$ is closer to the Nash equilibrium than $\pi^2$ in the transitive dimension, which means that $U(\pi^1,\pi^2) > 0$.
\end{itemize}
\end{proof}

\subsection{Proof of Theorem 3.5}

\begin{theorem}
    In symmetric zero-sum games, given the population $\mathcal{P}_{i} = \mathcal{P}_{j} = \{ \pi^1_i,\cdots,\pi^t_i \}$, $\forall \pi^k_i \in \mathcal{P}_i$, we have $\mathcal{V}(\pi^k_i) \leq \mathcal{V}(\theta_i \mid \mathcal{P}_i)$. $\theta_i$ is the equilibrium of the meta-game corresponding to the population $\mathcal{P}_i$.
\end{theorem}
\begin{proof}
The population $\mathcal{P}_i$ can be viewed as a subgame. Applying Theorem 3.4 for this subgame we get the following property:
\begin{equation}
    \forall \pi^k_i \in \mathcal{P}_i, \ \mathcal{V}(\pi^k_i \mid \mathcal{P}_i) \leq \mathcal{V}(\theta_i \mid \mathcal{P}_i).
\end{equation}

Then we the following derivation:
\begin{equation}
\begin{aligned}
\mathcal{V}(\pi^k_i) & = U(\pi^k_i,\operatorname{BR}(\pi^k_i)) = - U(\operatorname{BR}(\pi^k_i),\pi^k_i) \\
& \leq - U(\operatorname{BR^*}(\pi^k_i),\pi^k_i) = U(\pi^k_i,\operatorname{BR^*}(\pi^k_i)) \quad \text{where}  \   \operatorname{BR}^*(\pi^k_i) = \operatorname{argmin}_{\pi^\prime \in \mathcal{P}_i} U(\pi^k_i,\pi^\prime).  \\
& \leq U(\theta_i,\operatorname{BR^*}(\theta_i)) \\
& = \mathcal{V}(\theta_i \mid \mathcal{P}_i).
\end{aligned}
\end{equation}

From this theorem, we can see that $\mathcal{V}(\theta_i \mid \mathcal{P}_i)$ is an upper bound for $\mathcal{V}(\pi^k_i)$. This suggests that exploring new strategies in the neighborhood of $\theta_i$ increase the probability of improving the advantage of population.
\end{proof}


\subsection{Proof of Theorem 3.6}

\begin{theorem}
    Given the population $\mathcal{P}_{i} = \mathcal{P}_{j} = \{ \pi^1_i,\cdots,\pi^t_i \}$ and the equilibrium $\theta_i$, if $\mathcal{V}(\theta_i) < 0$, there exists $\Delta \pi_i \in \mathcal{A}$ and $\delta>0$ which satisfies:
    \begin{equation}
        \forall \ 0 < d < \delta, \quad \mathcal{V}( (1-d) \cdot \theta_i + d \cdot \Delta \pi_i ) >  \mathcal{V}(\theta_i),
    \end{equation}
\end{theorem}

\begin{proof}
From Theorem 3.4 we have that the inverse of advantage function $-\mathcal{V}(\pi_i)$ is convex about $\pi_i$. Since $\mathcal{V}(\theta_i) < 0 = \operatorname{max} \mathcal{V}(\pi_i)$, from the convexity of the function $-\mathcal{V}$ we can find a direction of descent in the domain of the strategy $\mathcal{D} = \{p_1,\cdots,p_m \mid p_i \geq 0, \sum p_i=1\}$:
\begin{equation}
    \exists \ \delta^\prime, \pi^\prime \in \mathcal{D}, \         \forall \ 0 < d < \delta^\prime, \quad \mathcal{V}( (1-d) \cdot \theta_i + d \cdot \pi^\prime ) >  \mathcal{V}(\theta_i).
\end{equation}
Since the domain of the strategy $\mathcal{D}$ is a convex combination of pure strategy space $\mathcal{A}$, $\pi^\prime \in \mathcal{D}$ can also be expressed as a convex combination of all elements in $\mathcal{A}$. This means that there exists $\Delta \pi_i \in \mathcal{A}$, which satisfies that $\langle \pi^\prime, \Delta \pi_i \rangle > 0$. We define that $\delta = \frac{\lvert \langle \pi^\prime, \Delta \pi_i \rangle \rvert}{\lvert \Delta \pi_i \rvert \cdot \lvert \pi^\prime \rvert} \cdot \delta^\prime$. Then we have:
\begin{equation}
    \forall \ 0 < d < \delta, \quad \mathcal{V}( (1-d) \cdot \theta_i + d \cdot \Delta \pi_i ) >  \mathcal{V}(\theta_i).
\end{equation}
\end{proof}

\subsection{Proof of Theorem 3.9}

\begin{theorem}
    In two-player general-sum simplified games, 
    \begin{itemize}
        \item $\forall i$, $\mathcal{V}_i(\pi_i)$ is Lipschitz continuous. 
        \item We assume that the joint strategy $(\pi_i,\pi_j)$ is a Nash equilibrium. If $\operatorname{BR}(\pi_i)  \cap \mathcal{A}_j = \{a_j^0\}$,  then $\mathcal{V}_i(\pi_i)$ is a local maximum of the advantage function $\mathcal{V}_i$.
        \item If joint strategies $(\pi^1_i,\pi^1_j)$ and $(\pi^2_i,\pi^2_j)$ are both NEs. Then $(\pi^1_i,\pi^1_j)$ Pareto dominates $(\pi^2_i,\pi^2_j)$ if and only if $\mathcal{V}(\pi^1_i) \geq \mathcal{V}(\pi^2_i)$ and $\mathcal{V}(\pi^1_j) \geq \mathcal{V}(\pi^2_j)$.
    \end{itemize}
\end{theorem}

\begin{proof}

\begin{itemize}
\item In Theorem 3.4, the proof of the Lipschitz continuity of the advantage function does not require that the game is zero-sum. Therefore, we can similarly prove that $\mathcal{V}_i(\pi_i)$ is Lipschitz continuous in two-player general-sum games. 

\item If the joint strategy $(\pi_i,\pi_j)$ is a Nash equilibrium, we have that $\pi_j \in \operatorname{BR}(\pi_i)$. We assume that:
\begin{equation}
    \operatorname{BR}(\pi_i) \cap \mathcal{A}_j = \{a^0_j\},
\end{equation}
which means that:
\begin{equation}
    \forall a^k_j \neq a^0_j, \ U_j(\pi_i,a^k_j) < U_j(\pi_i,a^0_j).
\end{equation}
From the proof of Theorem 3.4 we have that $\forall \pi_i, \forall a$, $U_i(\pi_i,a)$ is Lipschitz continuous about $\pi_i$. Then there must exsits $\delta >0$, which satisfies that:
\begin{equation}
    \forall \pi^\prime_i \in B_{\delta}(\pi_i), \ \left(\operatorname{BR}(\pi^\prime_i)  \cap \mathcal{A}_j \right) \subseteq \left( \operatorname{BR}(\pi_i) \cap \mathcal{A}_j \right),
\end{equation}
where $B_{\delta}(\pi_i)$ is the open ball of radius $\delta$ centered on $\pi_i$.
From Theorem 3.2, we have that $\{a^0_j\} = \operatorname{BR}(\pi^\prime_i)  \cap \mathcal{A}_j$, then we have:
\begin{equation}
\begin{aligned}
    \mathcal{V}_i(\pi_i) & = U(\pi_i,\pi_j) = U(\pi_i,a^0_j) \\
    & \geq U(\pi^\prime_i,\pi_j) = U(\pi^\prime_i,a^0_j) \quad (\text{becaue} (\pi_i,\pi_j) \text{is Nash equilibrium}) \\
    & = \mathcal{V}_i(\pi^\prime_i).
\end{aligned}
\end{equation}
This illustrates that $\mathcal{V}_i(\pi_i)$ is a local maximum of the advantage function $\mathcal{V}_i$.

\item $(\pi^1_i,\pi^1_j)$ Pareto dominate $(\pi^2_i,\pi^2_j)$ is equivalent to $U_i(\pi^1_i,\pi^1_j) \geq U_i(\pi^2_i,\pi^2_j)$ and $U_j(\pi^1_i,\pi^1_j) \geq U_j(\pi^2_i,\pi^2_j)$. Since $(\pi^1_i,\pi^1_j)$ and $(\pi^2_i,\pi^2_j)$ are both Nash equilibrium, this holds if and only if $\mathcal{V}(\pi^1_i) \geq \mathcal{V}(\pi^2_i)$ and $\mathcal{V}(\pi^1_j) \geq \mathcal{V}(\pi^2_j)$.
\end{itemize}
\end{proof}

\subsection{Proof of Theorem 3.10}

\begin{theorem}
    In two-player general-sum simplified games, the current population is $\mathcal{P}_i=\{\pi^1_i,\cdots,\pi^t_i\}$. $\theta^*_i$ is the global maximum in $\operatorname{hull}(\mathcal{P}_i)$. Then there must exists non-zero measure set $\mathcal{D}^\prime \subset \operatorname{hull}(\mathcal{P}_i)$, which satisfies that if $\theta^{\prime}_i$ is a local maximum in $\mathcal{D}^\prime$, then $\mathcal{V}_i(\theta^{\prime}_i) = \mathcal{V}_i(\theta^{*}_i)$.
\end{theorem}

\begin{proof}
We assume that the strategy of the player $i$ is $\pi_i=(p_1,\cdots,p_{\lvert \mathcal{A}_i \rvert})$. For $k \in \{1,\cdots,\lvert \mathcal{A}_j \rvert \}$, we define $g_k(\pi_i) = U_j(\pi_i, a_j^k)$, where $a_j^k \in \mathcal{A}_j$. Then we have:
\begin{equation}
    g_k(p_1,\cdots,p_{\lvert \mathcal{A}_i \rvert}) = p_1 U_j(a_i^1,a_j^k) + \cdots + p_{\lvert \mathcal{A}_i \rvert} U_j(a_i^{\lvert \mathcal{A}_i \rvert},a_j^k).
\end{equation}
If the elements in set $\operatorname{BR}(\pi_i) \cap \mathcal{A}_j$ are not unique, there must exists $k$ and $k^\prime$ which satisfies that $g_k(\pi_i) = g_{k^\prime}(\pi_i)$. Since the game is simplified, $\left(U_j(a_i^1,a_j^k),\cdots,U_j(a_i^{\lvert \mathcal{A}_i \rvert},a_j^k)\right)$ and $\left(U_j(a_i^1,a_j^{k^\prime}),\cdots,U_j(a_i^{\lvert \mathcal{A}_i \rvert},a_j^{k^\prime})\right)$ are linearly independent vectors. This indicates that $\pi_i$ satisfying $g_k(\pi_i) = g_{k^\prime}(\pi_i)$ is a zero measure set and non-dense in the domain $\Pi_i$. 

We define
\begin{equation}
    D^0 = \{\pi_i \mid \pi_i \in \operatorname{hull}(\mathcal{P}_i), (\operatorname{BR}(\pi_i) \cap \mathcal{A}_j) \ \text{is a singleton set}\}.
\end{equation}
Since $\pi_i$ satisfying that there exists $k$ and $k^\prime$ with $g_k(\pi_i) = g_{k^\prime}(\pi_i)$ is non-dense in the domain $\Pi_i$, we consider the projection of those $\pi_i$ onto $\operatorname{hull}(\mathcal{P}_i)$. This indicates that either $D^0$ is a empty set (which means that $\pi_i$ intersects different functions $g$ covers the plane of $\operatorname{hull}(\mathcal{P}_i)$), or $(\operatorname{hull}(\mathcal{P}_i) \setminus D^0)$ is a non-dense set. 

$D^0$ is a empty set means that $\forall \pi_i \in \operatorname{hull}(\mathcal{P}_i)$, $g_k(\pi_i) = g_{k^\prime}(\pi_i)$. Since the game is simplified, $\forall \pi_i \in \operatorname{hull}(\mathcal{P}_i)$, $U_i(\pi_i,a^k_j) = U_{i}(\pi_i,a^{k^\prime}_j)$. Thus, we can remove $a^{k^\prime}_j$ feom $\mathcal{A}_j$, which does not affect the calculation of $\mathcal{V}_i(\pi_i)$.

If $(\operatorname{hull}(\mathcal{P}_i) \setminus D^0)$ is a non-dense set, we consider separately whether $\theta^*_i \in D^0$. If $\theta^*_i \in D^0$, there must exists non-zero measure set $\mathcal{D}^1 \subseteq \operatorname{hull}(\mathcal{P}_i)$, which satisfies that:
\begin{equation}
    \forall \pi_i \in \mathcal{D}^1, \ \operatorname{BR}(\pi_i) \cap \mathcal{A}_j = \operatorname{BR}(\theta^*_i) \cap \mathcal{A}_j.
\end{equation}
This indicates that $\mathcal{V}_i(\pi_i)$ is a linear function about $\pi_i$. Since $\theta^*_i$ is the global maximum of this linear function, there must exists non-zero measure set $\mathcal{D}^\prime \subseteq \operatorname{hull}(\mathcal{P}_i)$, which satisfies that if $\theta^{\prime}_i$ is a local maximum in $\mathcal{D}^\prime$, then $\mathcal{V}_i(\theta^{\prime}_i) = \mathcal{V}_i(\theta^{*}_i)$.

 If $\theta^*_i \in (\operatorname{hull}(\mathcal{P}_i) \setminus D^0)$, we assume that:
\begin{equation}
    \operatorname{argmax}_k g_k(\theta^*_i) = \{1,\cdots,l\}.
\end{equation} 
 Due to the Lipschitz continuity of $U$, there must exists $d > 0$, which satisfies that:
 \begin{equation}
     \forall \Delta \pi_i,  \ \operatorname{BR}(\theta^*_i + \frac{\Delta \pi_i}{\lvert \Delta \pi_i \rvert} \cdot d) \subseteq \{\pi_j^1,\cdots,\pi_j^l\}.
 \end{equation}
Since $g_k(\pi_i)$ is a hyperplane corresponds to a linear function, the value of function $g_k(\pi_i)$ on the ray $\theta^*_i + \frac{\Delta \pi_i}{\lvert \Delta \pi_i \rvert} \cdot \delta, \ 0< \delta <d$ is either all maximal or all non-maximal for all $k$. Thus we can assume that there is a unique pure strategy best response $\pi_j^k, k \in \{ 1,\cdots,l\}$ for a strategy on that ray.
 
Thus, the open ball $B_{d}(\theta^*_i)$ can be divided into at most $\lvert \mathcal{A}_j \rvert$ region $D^k, k \in \{1,\cdots,\lvert \mathcal{A}_j \rvert\}$. In every region $D^k$, $\mathcal{V}_i(\pi_i)$ is linear function about $\pi_i$. Since $\theta^*_i$ is the global maximum of this linear function, there must exists non-zero measure set $\mathcal{D}^\prime \subseteq B_{d}(\theta^*_i)$, which satisfies that if $\theta^{\prime}_i$ is a local maximum in $\mathcal{D}^\prime$, then $\mathcal{V}_i(\theta^{\prime}_i) = \mathcal{V}_i(\theta^{*}_i)$.
\end{proof}


\section{Algorithm Introduction and the Pseudo-Code}

\subsection{Classic PSRO Algorithm}
\label{Classic PSRO Algorithm}
\begin{algorithm}[h]
\caption{Policy-Space Response Oracles}
\label{alg1}
\textbf{Input}: initial policy populations for all players $\mathcal{P}$. Compute the expected utilities $U^{\mathcal{P}}$ for each joint $\pi \in \mathcal{P}$. Initialize meta-strategies $\theta_i = \operatorname{UNIFORM}(\mathcal{P}_i)$  
\begin{algorithmic}[1] 
\WHILE{iters $e$ in $\{1,2,\cdots\}$}
\FOR{player $i \in \{1,\cdots,n\}$ }
\FOR{many episodes}
\STATE Sample $\pi_{-i} \sim \theta_{-i}$
\STATE Train oracle $\pi^\prime_i$ over $\mathcal{O}(\pi^\prime_i,\pi_{-i})$
\ENDFOR
\STATE $\mathcal{P}_i = \mathcal{P}_i \cup \{\pi^\prime_i\}$
\ENDFOR
\STATE Compute missing entries in $U^{\mathcal{P}}$ from $\mathcal{P}$
\STATE Compute a meta-strategy $\theta$ from $U^{\mathcal{P}}$
\ENDWHILE 
\end{algorithmic}
\textbf{Output}: Current solution strategy $\theta_i$ for player $i$.
\end{algorithm}

Pseudo-code of the classic PSRO algorithm is given in Algorithm \ref{alg1}. $\operatorname{UNIFORM}$ denotes random sampling according to the uniform distribution. The two main components of the algorithm are the exploration of the new strategy $\pi^\prime_i$ and the computation of the meta-strategy $\theta$. In this paper, we focus on improving the PSRO framework from the perspective of new strategy exploration. We use the meta-strategy solver with exactly the same parameters as the other PSRO algorithms in our comparison experiments \cite{10,2}.

\subsection{A-PSRO for Solving Zero-Sum Games}
\label{A-PSRO for Solving Zero-Sum Games1}
In this section, we provide the algorithm for the generation of new strategy, and the rest of the framework is the same as other PSRO algorithms. We assume that the current population is $(\mathcal{P}_i,\mathcal{P}_j)$, where $\mathcal{P}_i = \{\pi^1_i,\cdots,\pi^t_i\}$.

In our algorithm, the agents initially decide to update the last strategy $\pi^t_i$. The new strategy $\pi^{t+1}_i$ is generated and incorporated into the population only if the update process does not enhance the utility. As LookAhead updates the strategy in the transitive dimension, we set its learning rate lower than $\Vert \theta_i \Vert^{\infty}$ to prevent stagnation as the strategy approaches the Nash equilibrium.

\begin{algorithm}[h]
\caption{Population updating process of A-PSRO in zero-sum games}
\label{alg2}
\textbf{Input}: Population $(\mathcal{P}_i,\mathcal{P}_j)$, meta-Nash equilibrium $(\theta_i, \theta_j)$, strategy to be updated of the agent $\pi^t_i$. \\
\textbf{Parameter}: diversity weight $\lambda_d$, learning rate $l_r$, improvement bound $c_m$. 
\begin{algorithmic}[1] 
\STATE Randomly generate $d_r \sim \mathbf{U}[0,1]$.
\IF{$d_r \leq \lambda_d$}
\STATE $\Delta \pi^t =\operatorname{argmax}_{\Delta \pi \in \mathcal{A}} \ [ \mathrm{EC}( \mathcal{P}_i \setminus \{\pi^t_i\} \cup\left\{(1-l_r) \cdot \pi^t_i + l_r \cdot \Delta \pi \right \} \mid \mathcal{P}_j ) ].$
\ELSE 
\STATE Randomly generate $l_r \sim \mathbf{U}[0, \text{min}(l_r,  \Vert \theta_i \Vert^{\infty})]$. 
\STATE $\Delta \pi^t = \operatorname{argmax}_{\Delta \pi \in \mathcal{A}} \ \mathcal{V}[ (1-l_r) \cdot \pi^t_i + l_r \cdot \Delta \pi  ]$.
\ENDIF
\STATE $\pi^{*}_i =  (1-l_r) \cdot \pi^t_i + l_r \cdot \Delta \pi $.
\IF {$\frac{\pi^{*}_i \times \mathcal{U}_i \times \theta_j}{\pi^{t}_i \times \mathcal{U}_i \times \theta_j} - 1 \geq c_m $}
\STATE $\pi^t_i = \pi^{*}_i$.
\ELSE
\STATE $\pi^t_i = \pi^{*}_i$. Then randomly generate $\pi^{t+1}_i$, $\mathcal{P}_i = \mathcal{P}_i \cup \{\pi^{t+1}_i\}$.
\ENDIF
\STATE \textbf{return} $\mathcal{P}_i$.
\end{algorithmic}
\textbf{Output}: $\mathcal{P}_i$
\end{algorithm}

In Algorithm \ref{alg2}, we combine the diversity module and our LookAhead module. The EC (expected cardinality) function is the diversity measure used in \cite{10}:
\begin{equation}
\begin{aligned}
    & \operatorname{EC}(\mathcal{P}_i \mid \mathcal{P}_j) := \operatorname{Tr} (\mathbf{I}-(\mathcal{L}+\mathbf{I})^{-1}) \\
    & \mathcal{L} = \mathcal{M}_i \mathcal{M}^T_i, \ \mathcal{M}_i =  \mathcal{P}_i \times U_i \times \mathcal{P}_j.
\end{aligned}
\end{equation}

In all zero-sum game experiments, we control the proportion of diversity and LookAhead modules with a uniform parameter $\lambda_d$. We find from our experimental results that A-PSRO achieves good convergences in games with different transitive and cyclic structures.

\subsection{A-PSRO for Solving Two-Player General-Sum Games}
\label{A-PSRO for Solving Two-Player General-Sum Games1}
In experiments with general-sum games, we find that the diversity module does not contribute to improving the reward of the strategy learning process. Therefore, the strategy exploration process of our algorithm A-PSRO contains only the LookAhead module. The Pseudo-Code of A-PSRO is given in Algorithm \ref{alg3}. In Algorithm \ref{alg3}, the meta-solver of the oracle $\mathcal{O}(\mathcal{P}_i,\mathcal{P}_j \mid \pi^k_i)$ is the fictitous play with 1000 itereations. In our experiments, we set the number of repeats $k=10$. The rest of the A-PSRO algorithm in the general-sum game is consistent with the zero-sum game.

\begin{algorithm}[h]
\caption{Population updating process of A-PSRO in general-sum games}
\label{alg3}
\textbf{Input}: Population $(\mathcal{P}_i,\mathcal{P}_j)$, meta-Nash equilibrium $(\theta_i, \theta_j)$, strategy to be updated of the agent $\pi^t_i$. \\
\textbf{Parameter}:learning rate $l_r$, improvement bound $c_m$. 
\begin{algorithmic}[1] 
\FOR{repeats $k$ in $\{1,2,\cdots\}$}
\STATE $\pi_i^k = \operatorname{UNIFORM}[\operatorname{hull}(\mathcal{P}_i)]$
\STATE $(\theta^k_i,\theta^k_j) = \mathcal{O}(\mathcal{P}_i,\mathcal{P}_j \mid \pi^k_i)$
\ENDFOR
\STATE $\theta_i = \operatorname{argmax}_{k} \mathcal{V}_i(\theta^k_i)$
\STATE Randomly generate $l_r \sim \mathbf{U}[0, \text{min}(l_r,  \Vert \theta_i \Vert^{\infty})]$. 
\STATE $\Delta \pi^t = \operatorname{argmax}_{\Delta \pi \in \mathcal{A}} \ \mathcal{V}_i[ (1-l_r) \cdot \pi^t_i + l_r \cdot \Delta \pi  ]$.
\STATE $\pi^{*}_i =  (1-l_r) \cdot \pi^t_i + l_r \cdot \Delta \pi $.
\IF {$\frac{\pi^{*}_i \times \mathcal{U}_i \times \theta_j}{\pi^{t}_i \times \mathcal{U}_i \times \theta_j} - 1 \geq c_m $}
\STATE $\pi^t_i = \pi^{*}_i$.
\ELSE
\STATE $\pi^t_i = \pi^{*}_i$. Then randomly generate $\pi^{t+1}_i$, $\mathcal{P}_i = \mathcal{P}_i \cup \{\pi^{t+1}_i\}$.
\ENDIF
\STATE \textbf{return} $\mathcal{P}_i$.
\end{algorithmic}
\textbf{Output}: $\mathcal{P}_i$
\end{algorithm}

\subsection{A-PSRO for Solving Multi-Player Games}
\label{A-PSRO for Solving Multi-Player Games1}

The main modification in applying A-PSRO algorithm to solve multi-player games is the computation of the advantage function. Unlike the two-player game with direct utilization of the best response $\operatorname{BR}$, the computation of the advantege of the strategy $\pi_i$ requires the usage of oracle $\mathcal{O}(\Pi_{-i} \mid \pi_i)$.

In multi-player zero-sum games, we adopt the joint best response as an approximation to the pessimistic equilibrium. The Pseudo-Code of calculating the advantage in A-PSRO is given in Algorithm \ref{alg4}.

\begin{algorithm}[h]
\caption{Calculation of the advantage function in multi-player zero-sum games}
\label{alg4}
\textbf{Input}: Strategy of the player $\pi_i$. 
\begin{algorithmic}[1] 
\STATE The identifying numbers set of other agents is $\{-i\} = \{1,\cdots,k\}$
\STATE The joint pure strategy space of other agents is $\mathcal{A}_{-i} = \mathcal{A}_{1} \times \cdots \times \mathcal{A}_{k}$
\FOR{$(a^m_1,\cdots,a^m_k) \in \mathcal{A}_{-i}$}
\STATE $u_m = U_i(\pi_i, \pi_{-i}=(a^m_1,\cdots,a^m_k))$
\ENDFOR
\STATE $\mathcal{V}_i(\pi_i) = \operatorname{argmin}_m u_m$ 
\end{algorithmic}
\textbf{Output}: $\mathcal{V}_i(\pi_i)$
\end{algorithm}
When the payment matrix of the game is given, the joint best response can be obtained by a simple computational procedure. Our experiment result \ref{fg6_1} shows the effectiveness of this approximation for the advantage function. In multi-player general-sum games, since the objective of agents maximizing rewards are independent, there is no need to specifically use pessimistic equilibrium oracle to compute the advantage. In our experiments, we directly use the fictitious play oracle with 1000 iterations to solve the advantage in multi-player general-sum games.

Besides the computation of the advantage function, A-PSRO is consistent with the two-player game in the multi-player game. The detailed framework of A-PSRO will be published along with the code in the future.


\section{Experiment Details and Additional Experiment Results}

\subsection{A-PSRO for Solving Zero-Sum Games}
\label{A-PSRO for Solving Zero-Sum Games2}
The parameter setting of zero-sum games is given in Table \ref{table1}. All experiments in this paper were run with CPU support on model Intel Core i9-10900KF CPU @ 3.70GHz. Experiments can be performed under both Windows and Linux systems. All source code required for conducting and analyzing the experiments will be made publicly available upon publication of the paper with a license that allows free usage for research purposes.

\begin{figure}[h]
\centering
\subfigure[AlphaStar] {\label{AlphaStar}
\includegraphics[width=0.22\columnwidth]{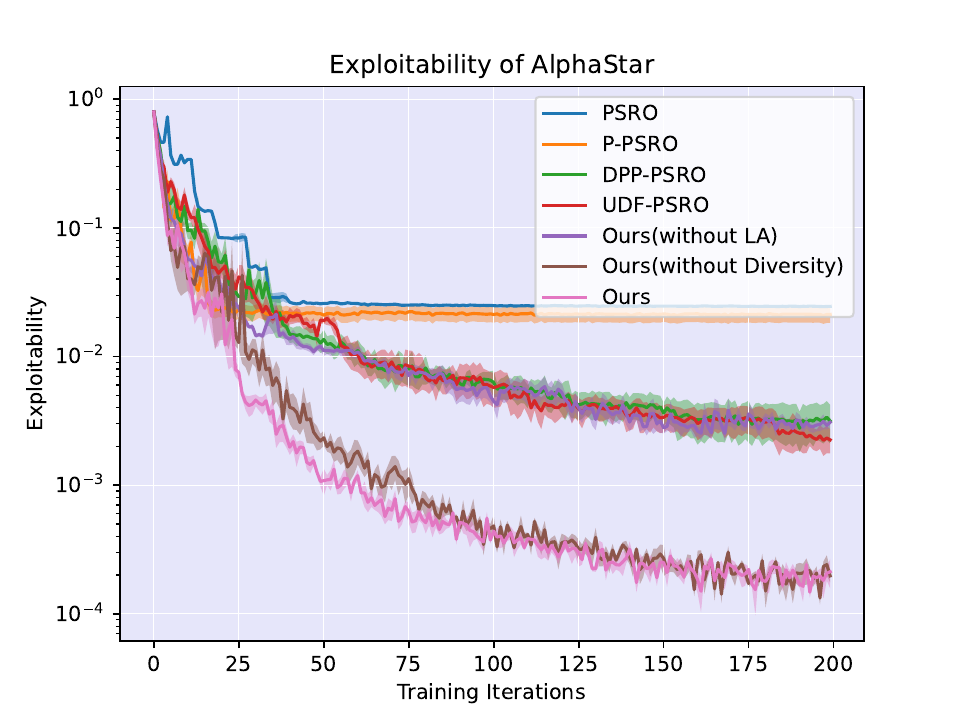}
}
\subfigure[Blotto] {\label{Blotto}
\includegraphics[width=0.22\columnwidth]{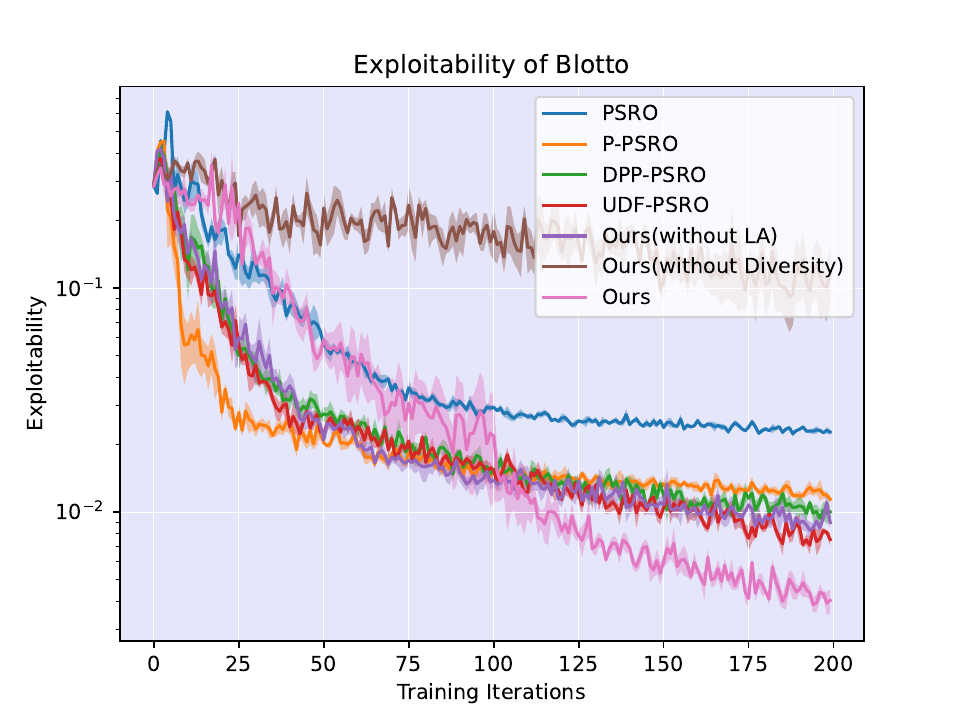}
}
\subfigure[Elo game] {\label{Elo game}
\includegraphics[width=0.22\columnwidth]{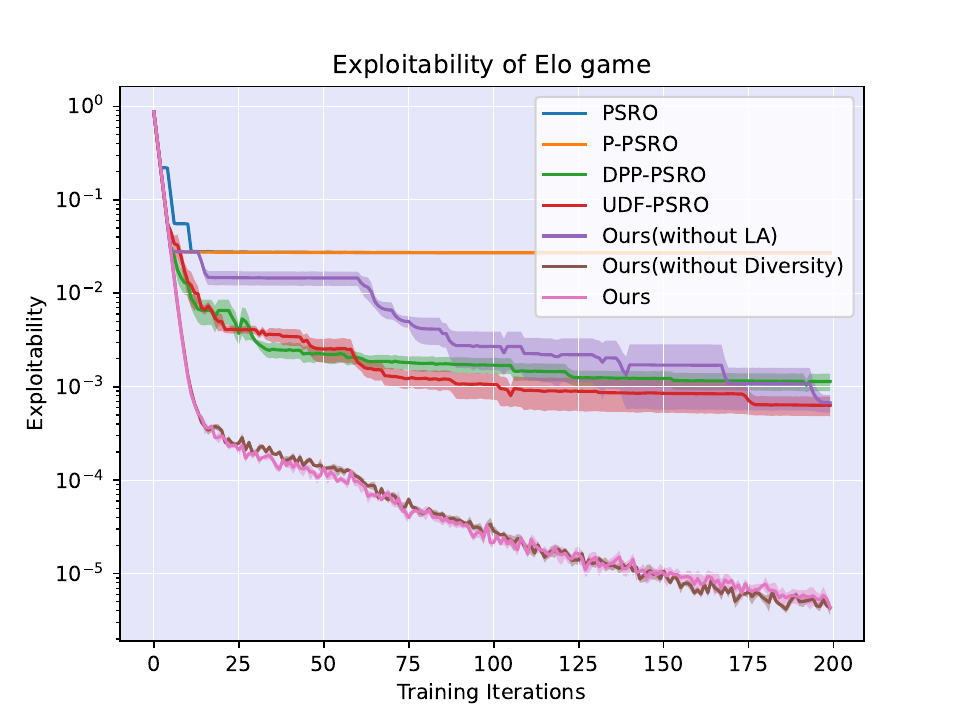}
}
\subfigure[Disc game] {\label{Disc game}
\includegraphics[width=0.22\columnwidth]{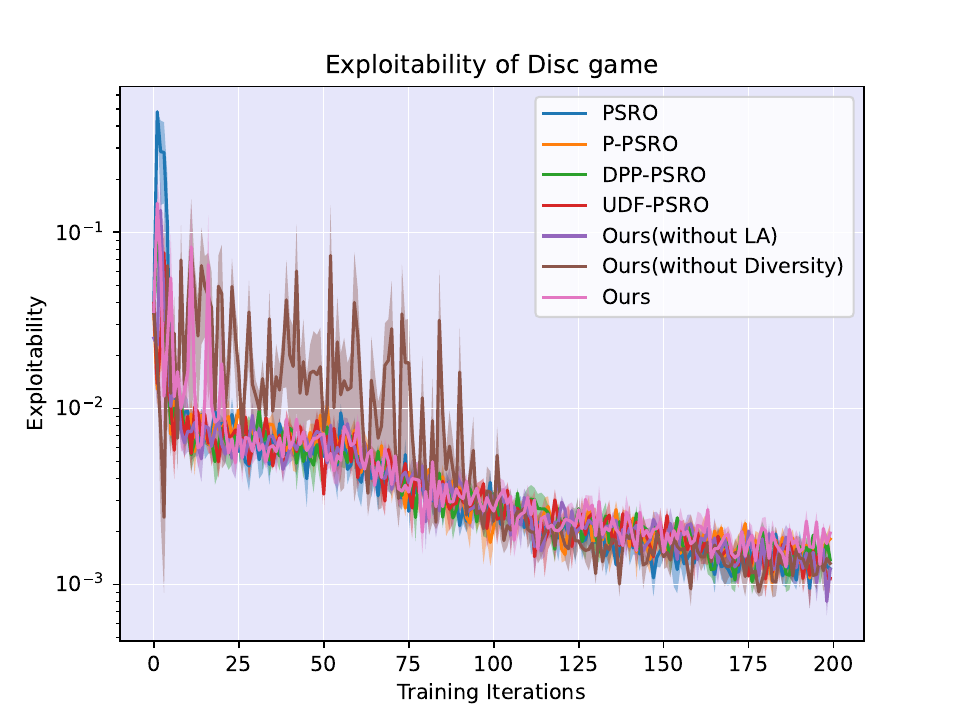}
}
\subfigure[Transitive game] {\label{Transitive game}
\includegraphics[width=0.22\columnwidth]{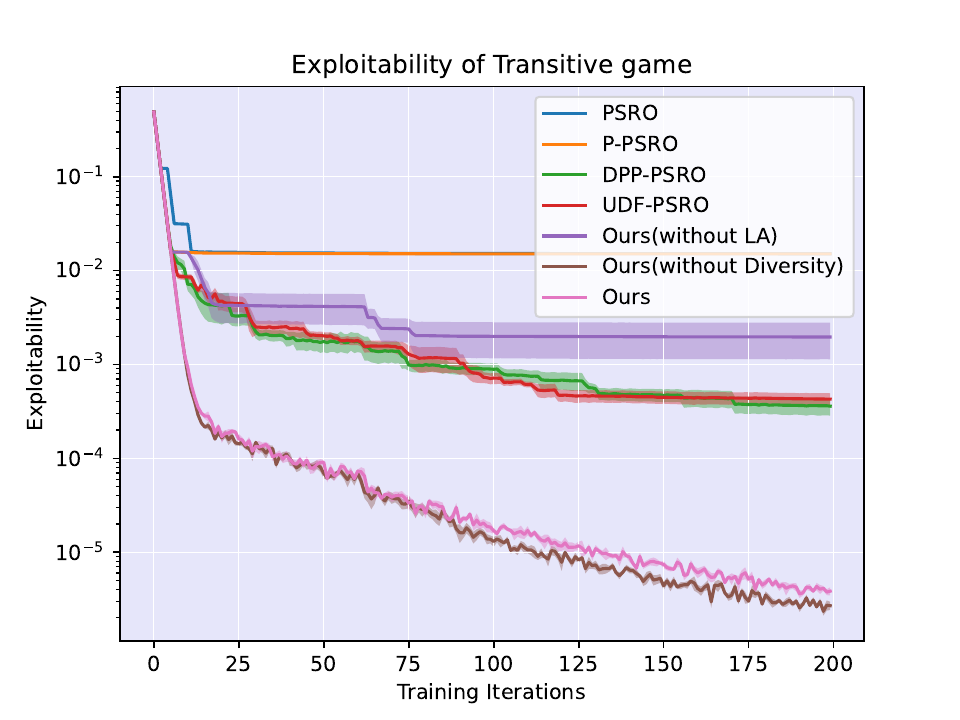}
}
\subfigure[Triangular game] {\label{Triangular game}
\includegraphics[width=0.22\columnwidth]{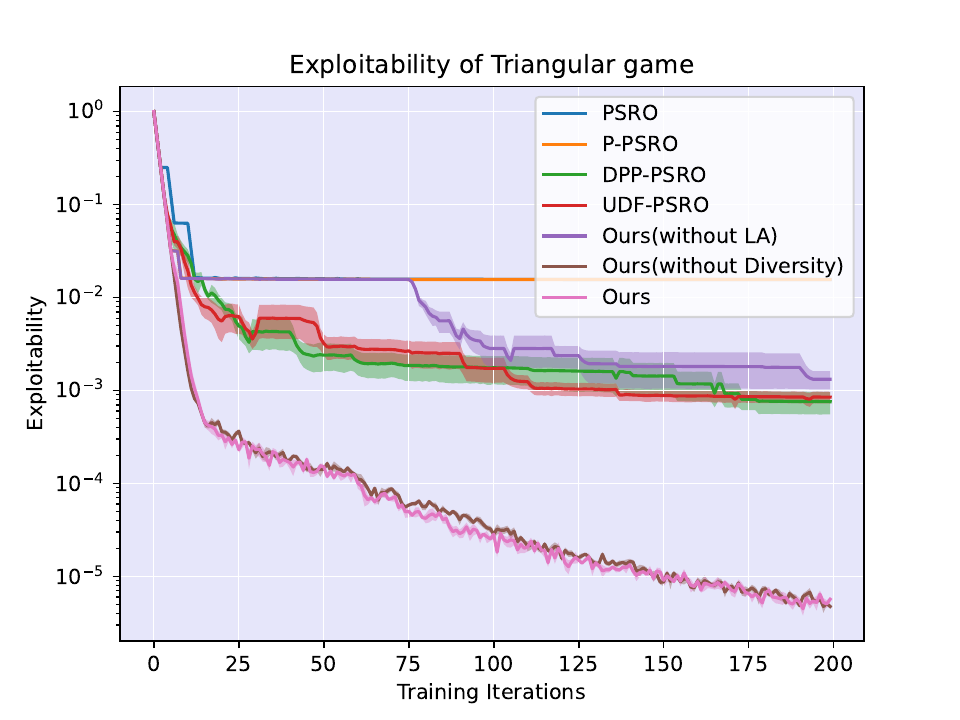}
}
\subfigure[Random game of skill] {\label{Random game of skill}
\includegraphics[width=0.22\columnwidth]{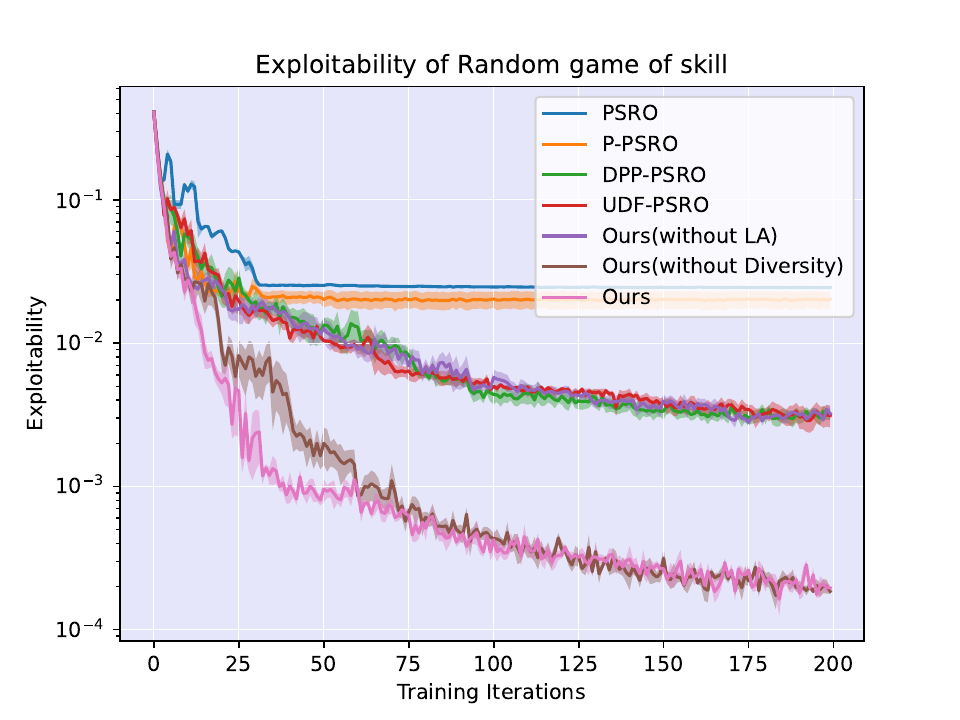}
}
\subfigure[Elo game + noise=0.1] {\label{Elo game + noise=0.1}
\includegraphics[width=0.22\columnwidth]{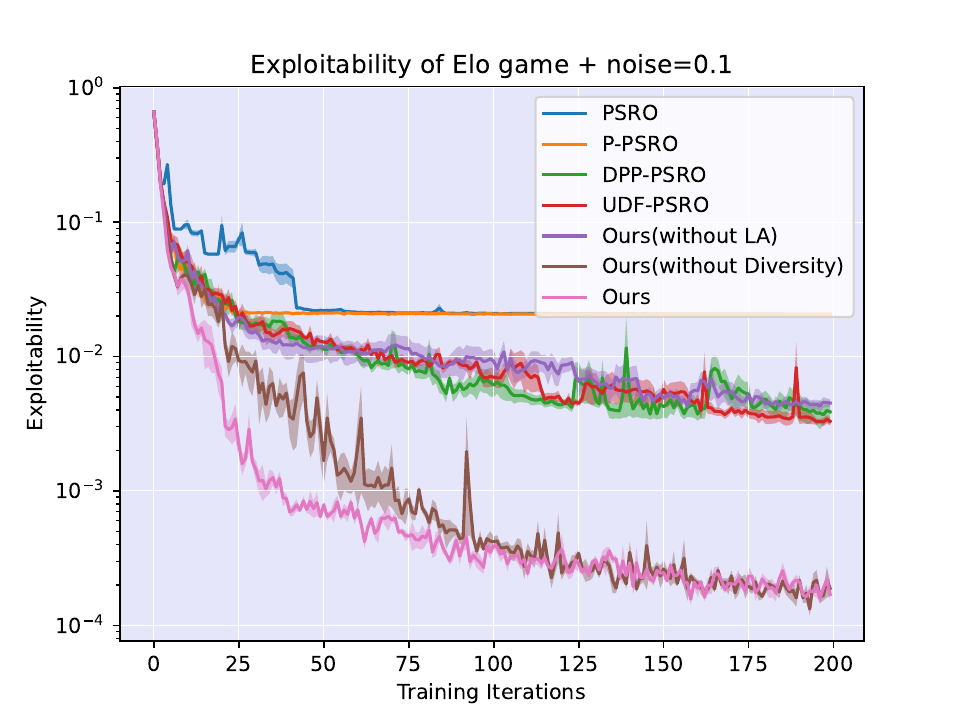}
}
\subfigure[Elo game + noise=0.5] {\label{Elo game + noise=0.5}
\includegraphics[width=0.22\columnwidth]{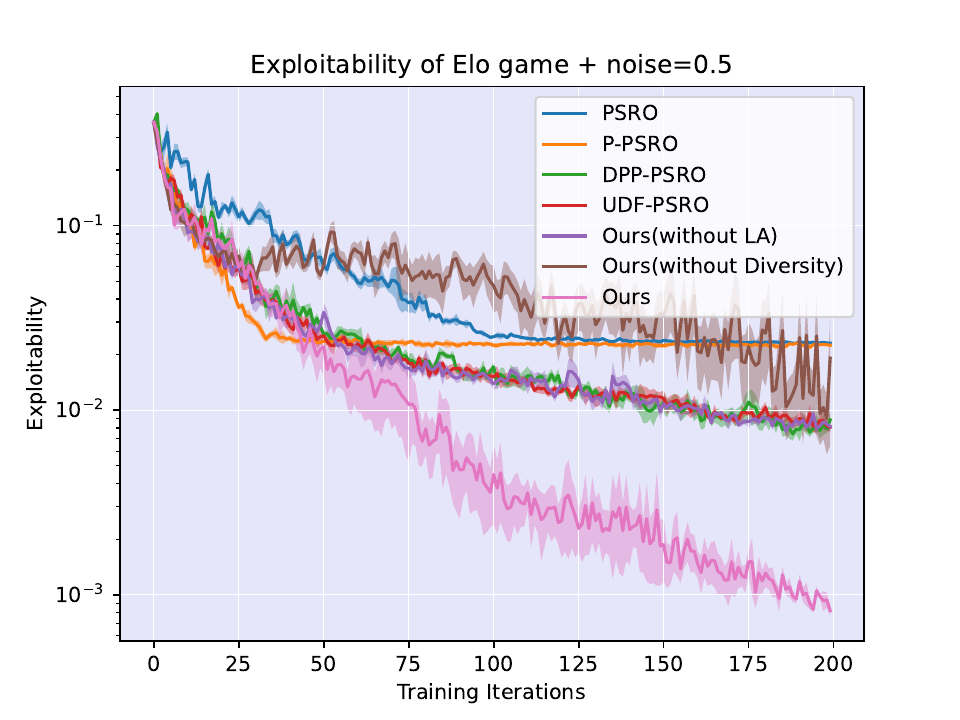}
}
\subfigure[Elo game + noise=1.0] {\label{Elo game + noise=1.0}
\includegraphics[width=0.22\columnwidth]{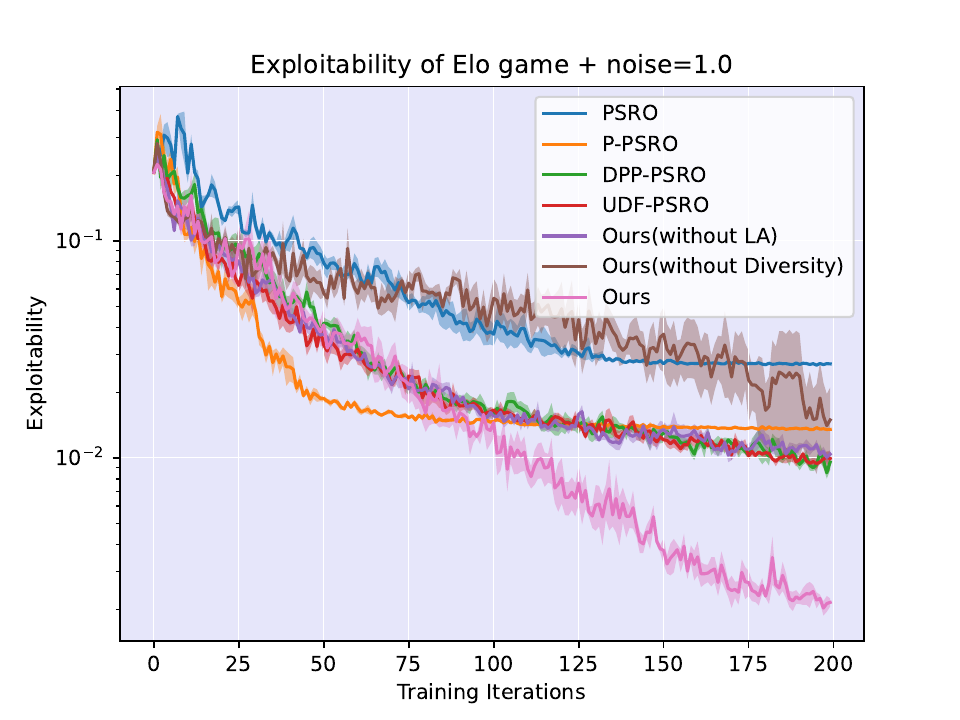}
}
\subfigure[Tic tac toe] {\label{tic tac toe}
\includegraphics[width=0.22\columnwidth]{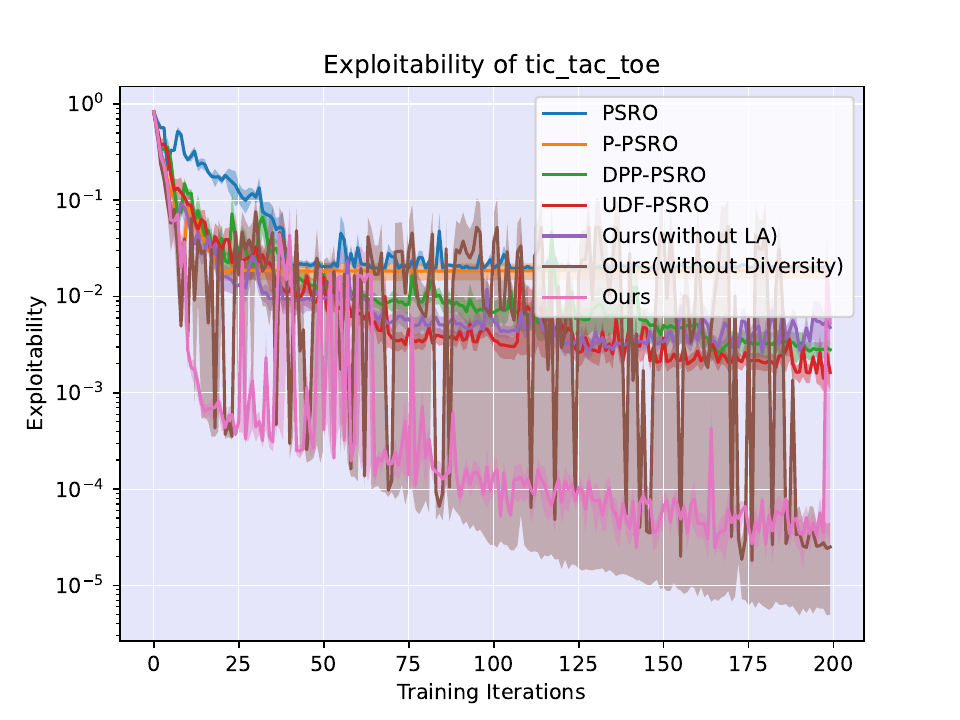}
}
\subfigure[Kuhn poker] {\label{Kuhn poker}
\includegraphics[width=0.22\columnwidth]{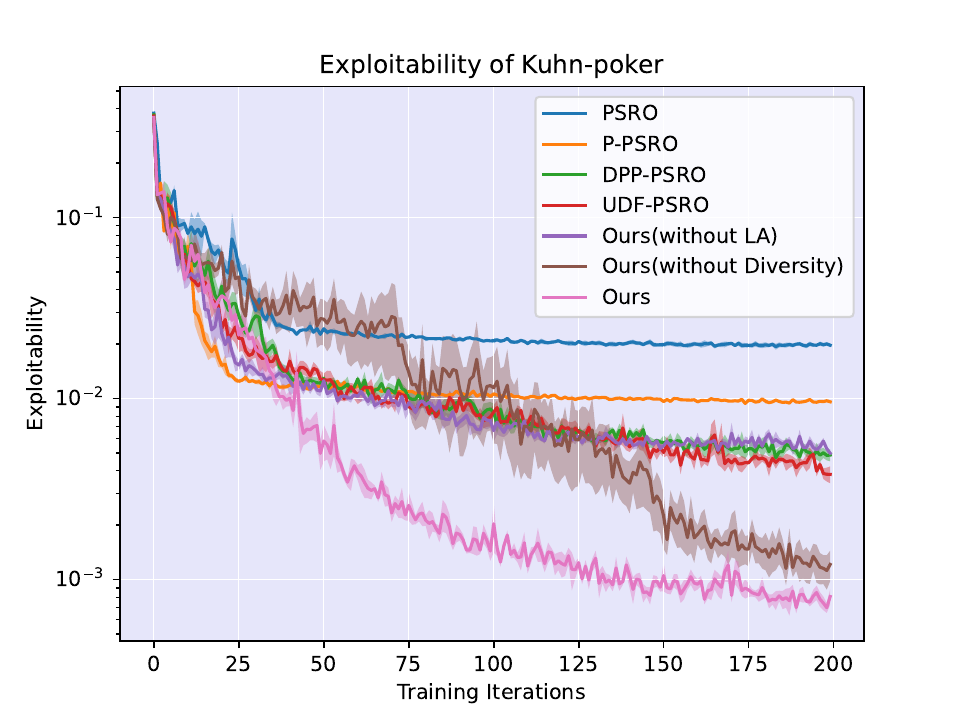}
}
\subfigure[Connect four] {\label{Connect four}
\includegraphics[width=0.22\columnwidth]{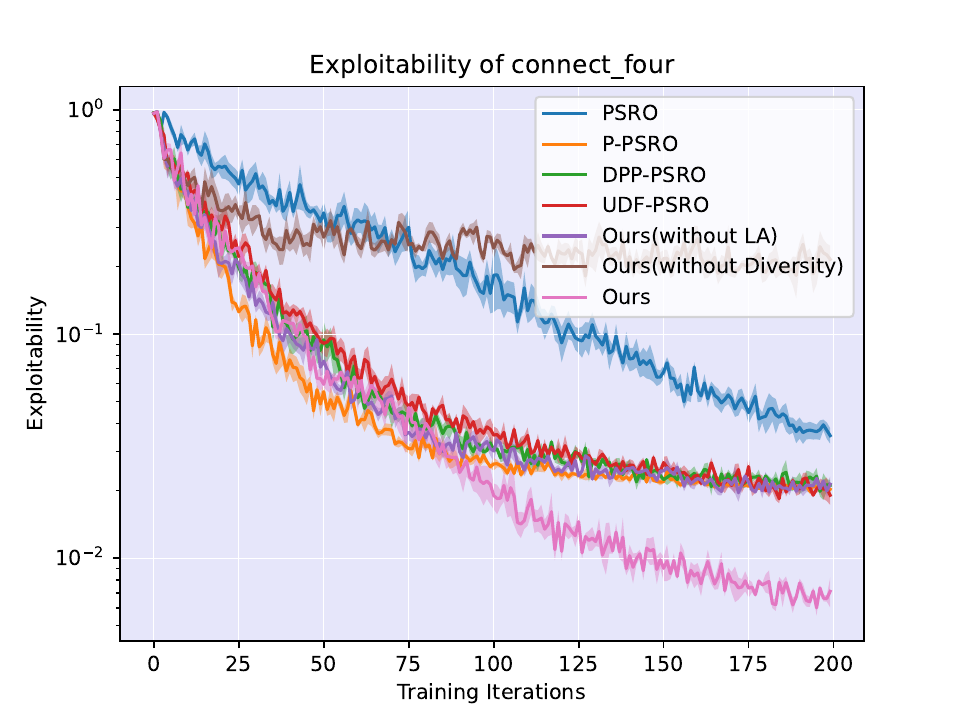}
}
\subfigure[3-move parity game] {\label{3-move parity game}
\includegraphics[width=0.22\columnwidth]{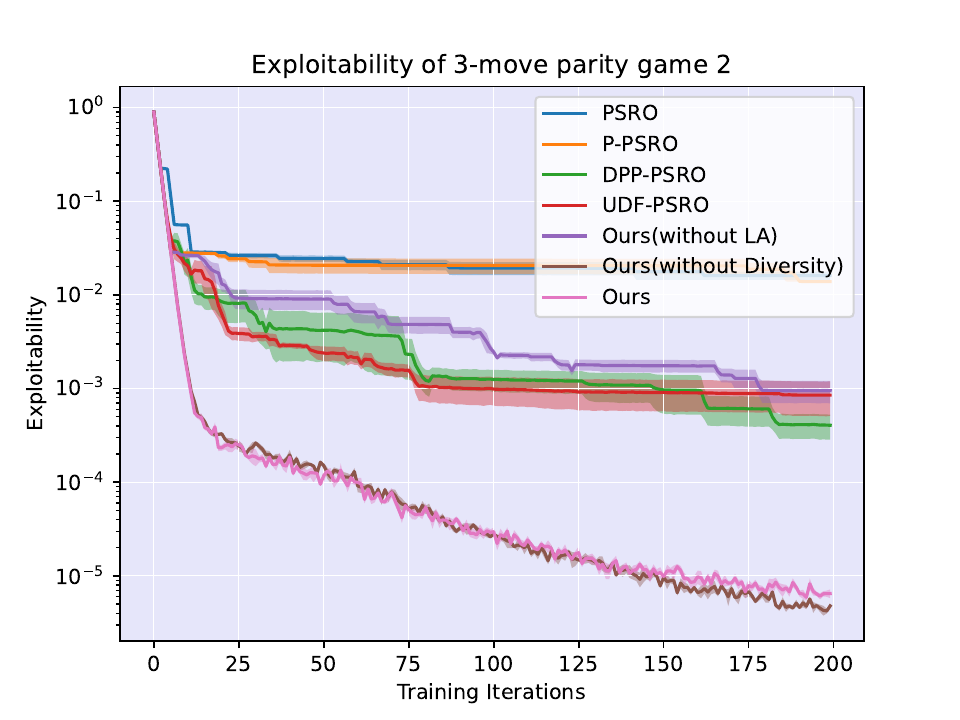}
}
\subfigure[Go (size=3)] {\label{Simplified Go game (size=3)}
\includegraphics[width=0.22\columnwidth]{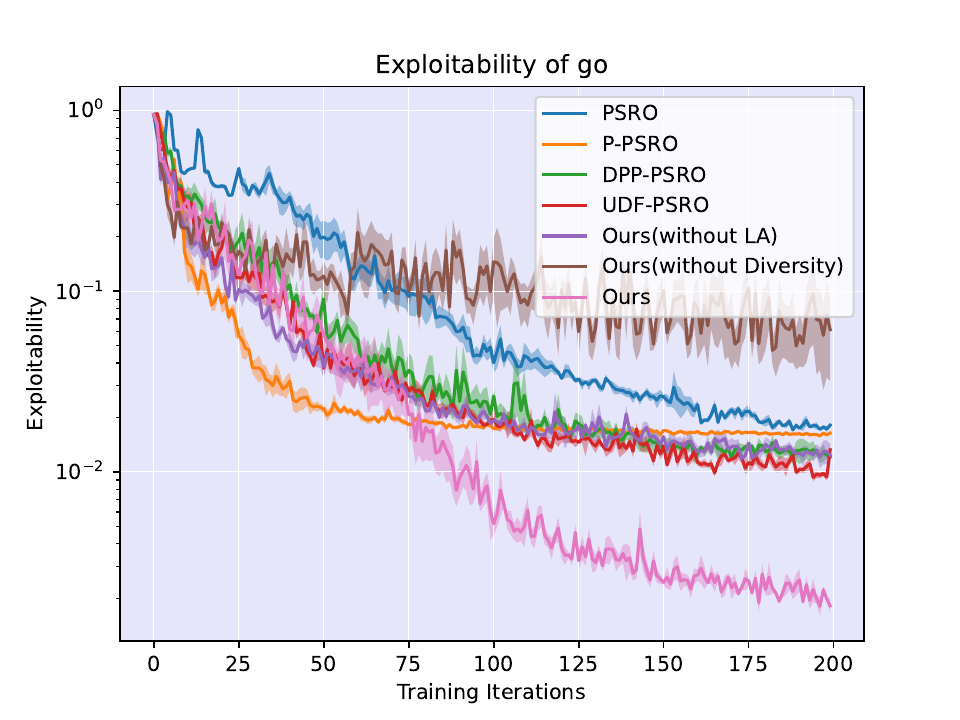}
}
\subfigure[Go (size=4)] {\label{Simplified Go game (size=4)}
\includegraphics[width=0.22\columnwidth]{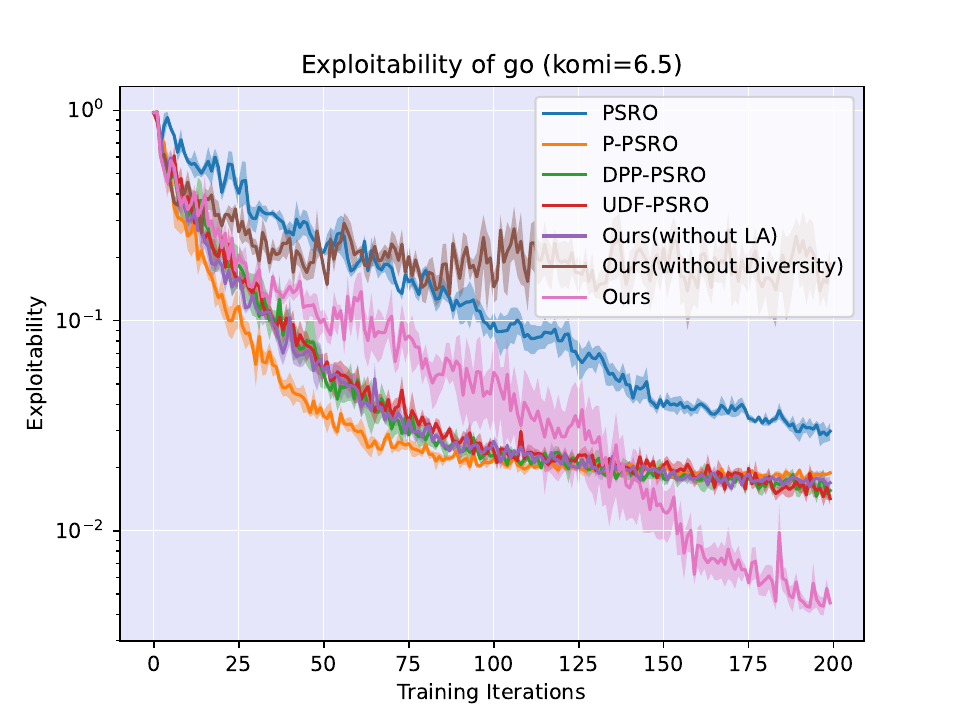}
}
\caption{The exploitability of the joint strategy learned by agents in various zero-sum games is depicted. The reduction in exploitability through population iterations can serve as an indicator of the effectiveness in approximating the Nash equilibrium.}
\label{supply1}
\end{figure}

\begin{figure}[h]
\centering
\subfigure[AlphaStar] {\label{AlphaStar2}
\includegraphics[width=0.22\columnwidth]{Ex_2_1.pdf}
}
\subfigure[Blotto] {\label{Blotto2}
\includegraphics[width=0.22\columnwidth]{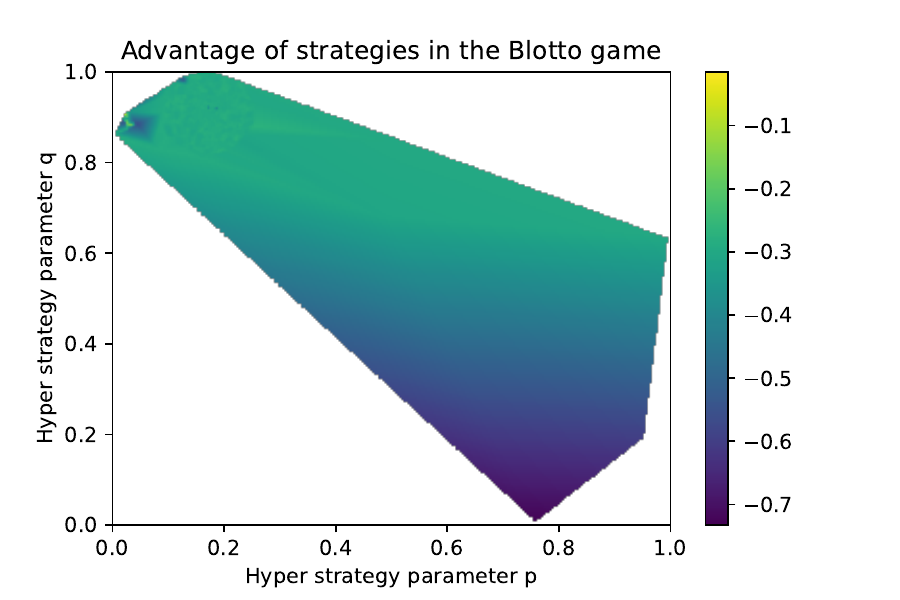}
}
\subfigure[Elo game] {\label{Elo game2}
\includegraphics[width=0.22\columnwidth]{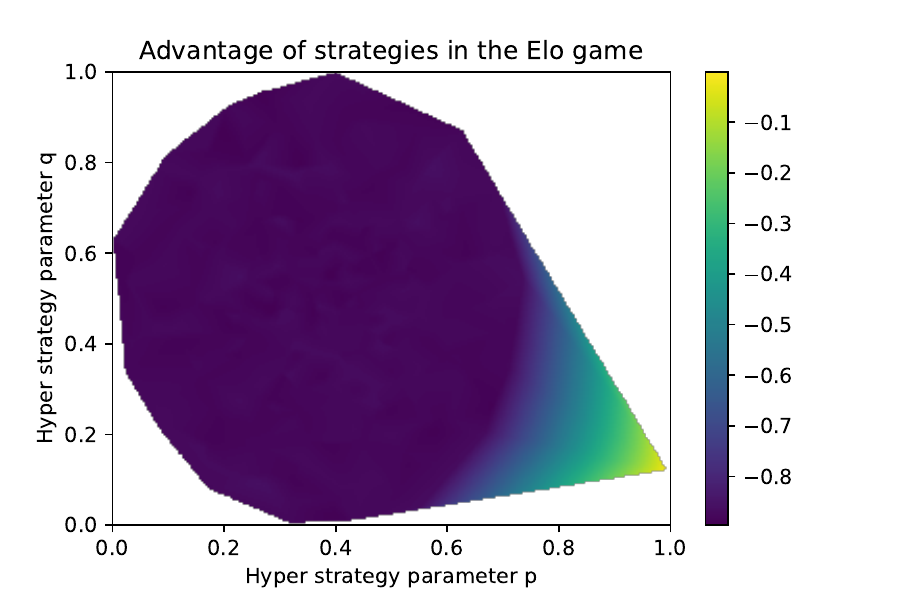}
}
\subfigure[Disc game] {\label{Disc game2}
\includegraphics[width=0.22\columnwidth]{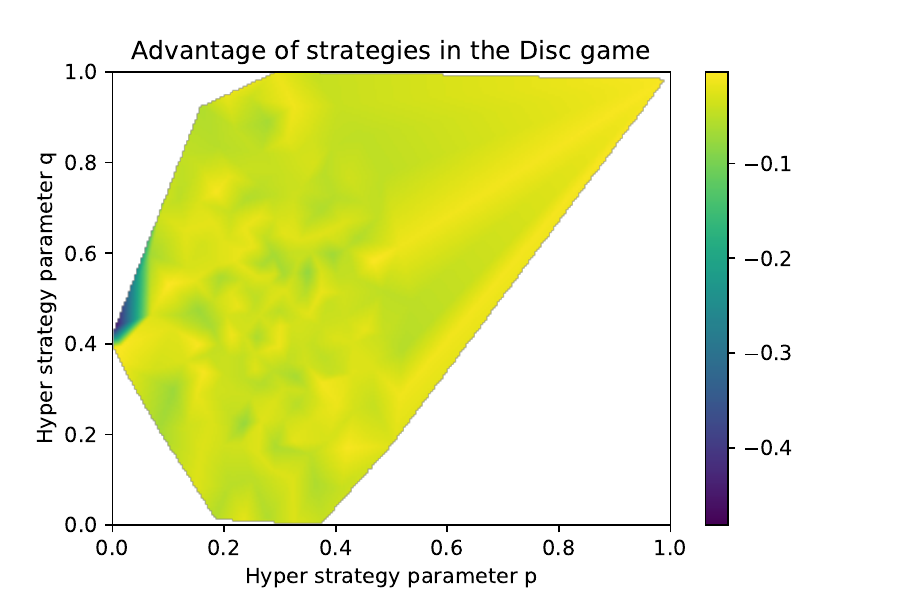}
}
\subfigure[Transitive game] {\label{Transitive game2}
\includegraphics[width=0.22\columnwidth]{Ex_2_2.pdf}
}
\subfigure[Triangular game] {\label{Triangular game2}
\includegraphics[width=0.22\columnwidth]{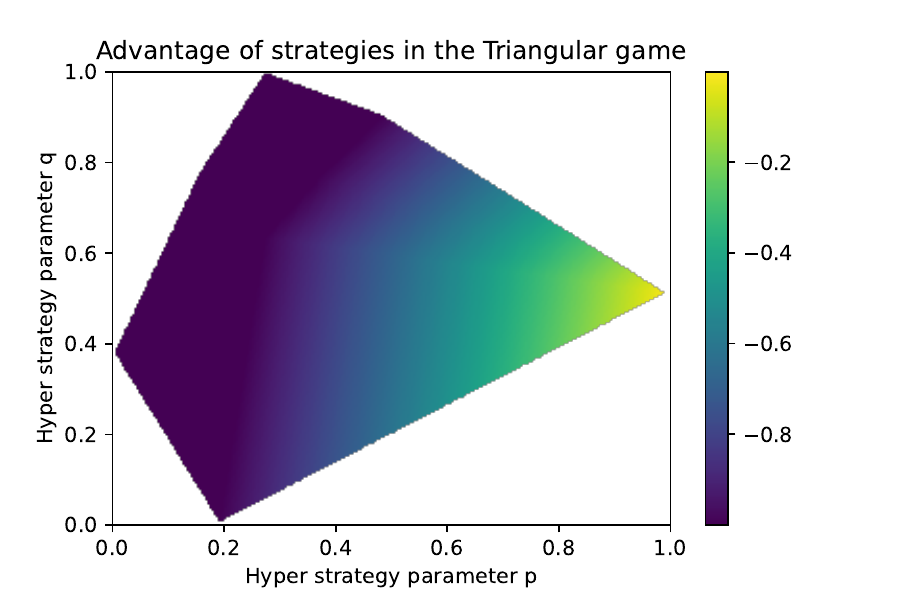}
}
\subfigure[Random game of skill] {\label{Random game of skill2}
\includegraphics[width=0.22\columnwidth]{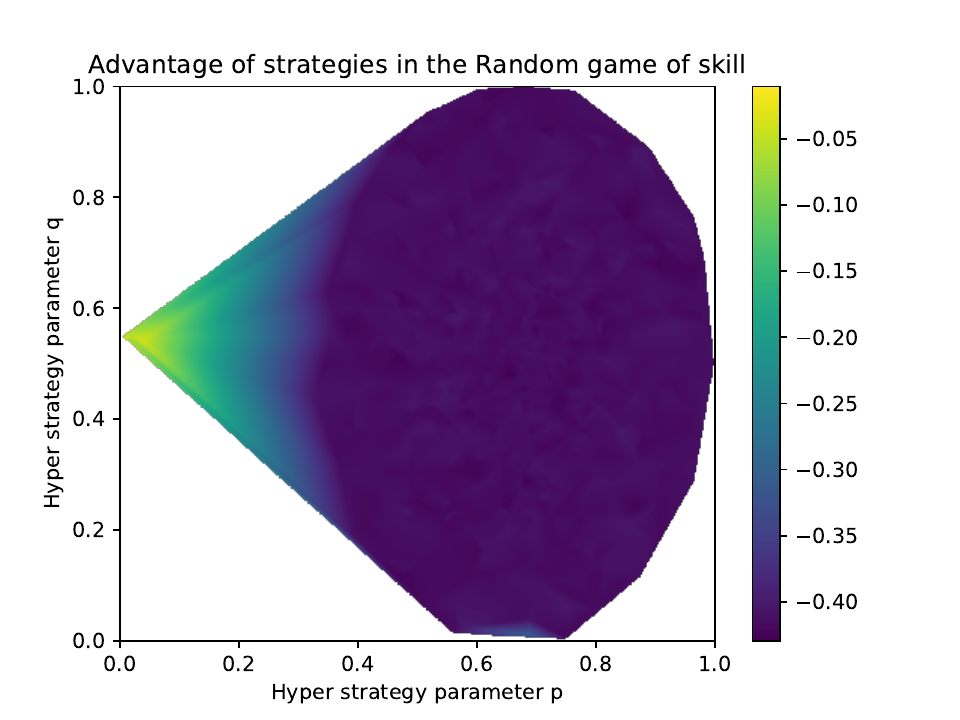}
}
\subfigure[Elo game + noise=0.1] {\label{Elo game + noise=0.12}
\includegraphics[width=0.22\columnwidth]{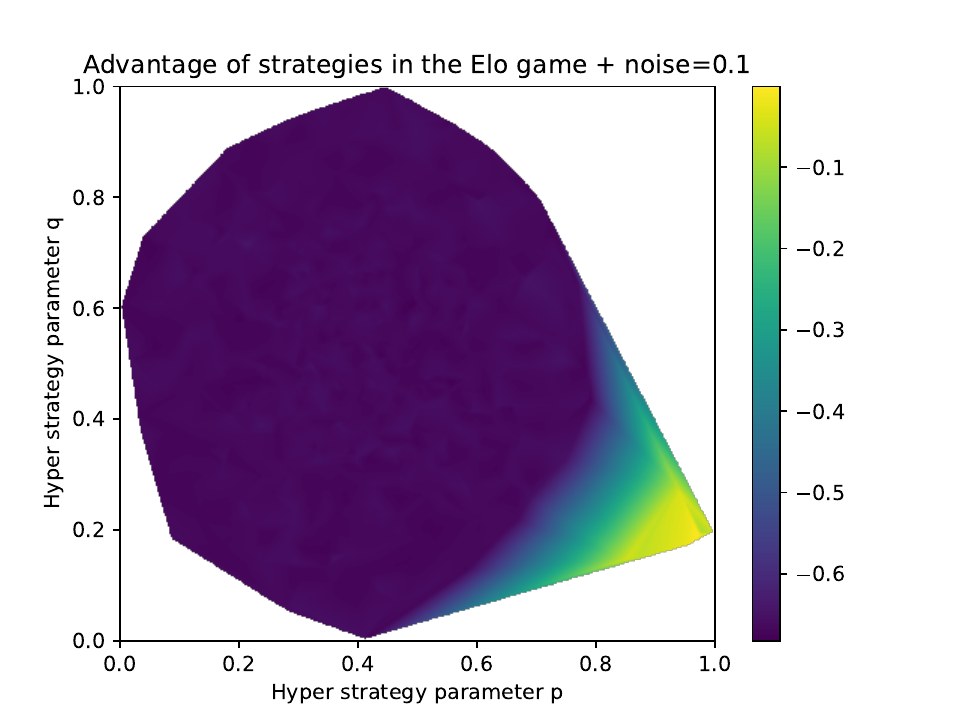}
}
\subfigure[Elo game + noise=0.5] {\label{Elo game + noise=0.52}
\includegraphics[width=0.22\columnwidth]{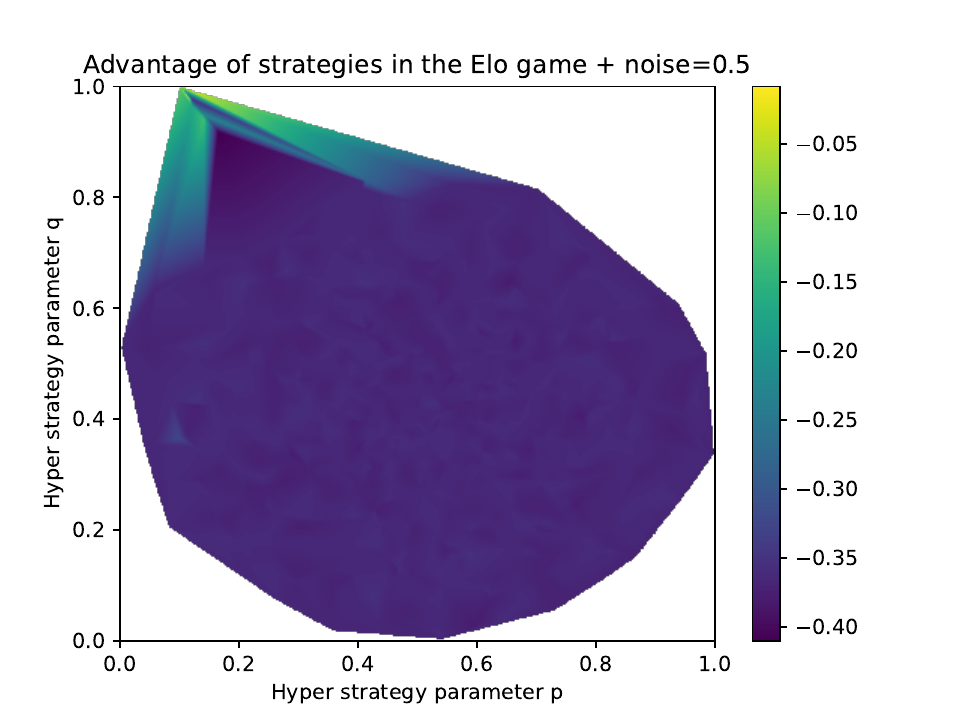}
}
\subfigure[Elo game + noise=1.0] {\label{Elo game + noise=1.02}
\includegraphics[width=0.22\columnwidth]{Ex_2_3.pdf}
}
\subfigure[Tic tac toe] {\label{tic tac toe2}
\includegraphics[width=0.22\columnwidth]{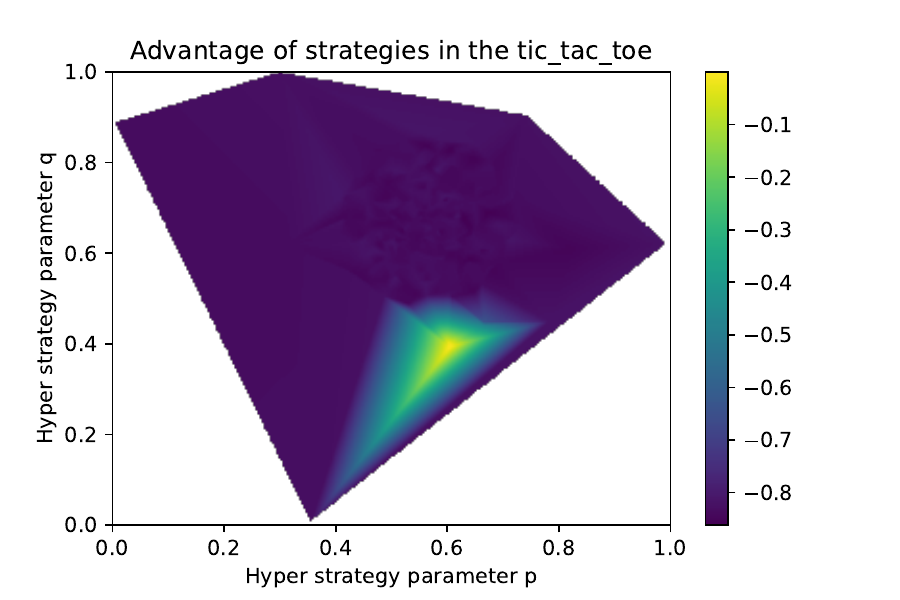}
}
\subfigure[Kuhn poker] {\label{Kuhn poker2}
\includegraphics[width=0.22\columnwidth]{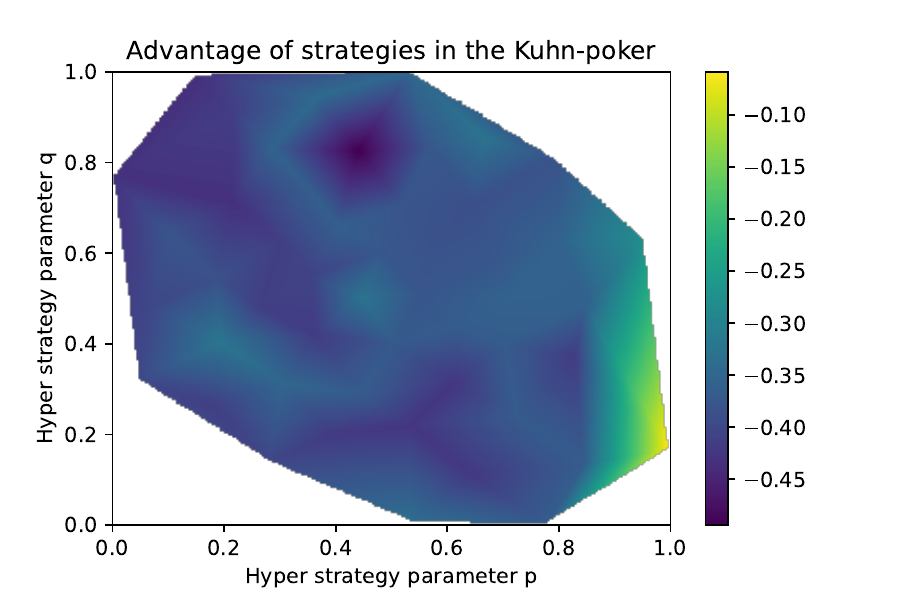}
}
\subfigure[Connect four] {\label{Connect four2}
\includegraphics[width=0.22\columnwidth]{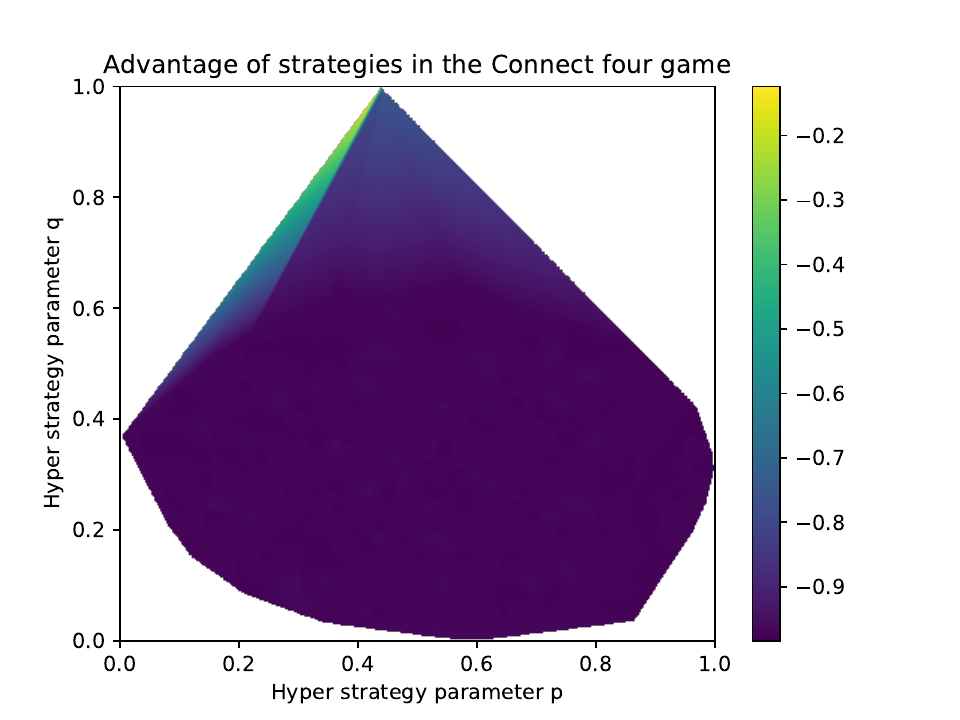}
}
\subfigure[3-move parity game] {\label{3-move parity game2}
\includegraphics[width=0.22\columnwidth]{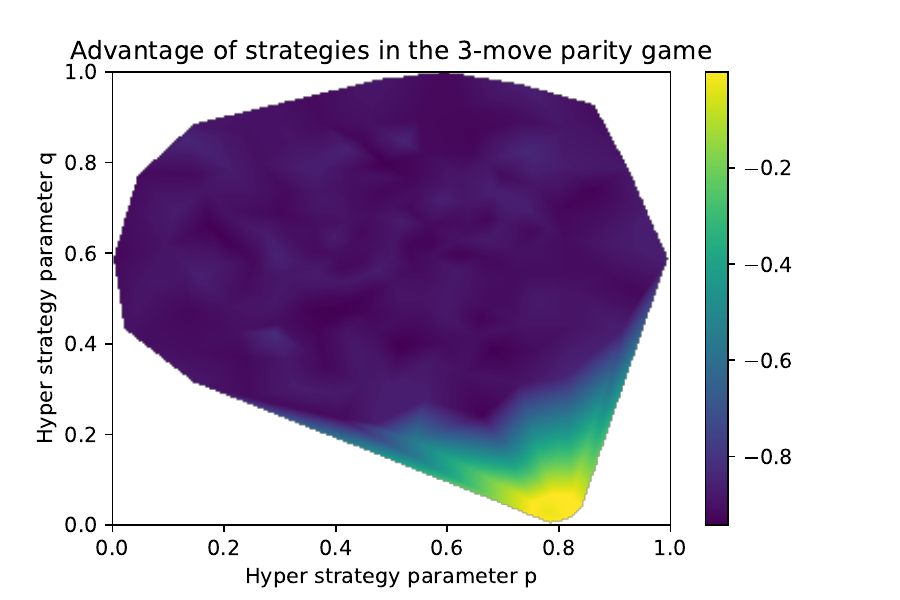}
}
\subfigure[Go (size=3)] {\label{Simplified Go game (size=3)2}
\includegraphics[width=0.22\columnwidth]{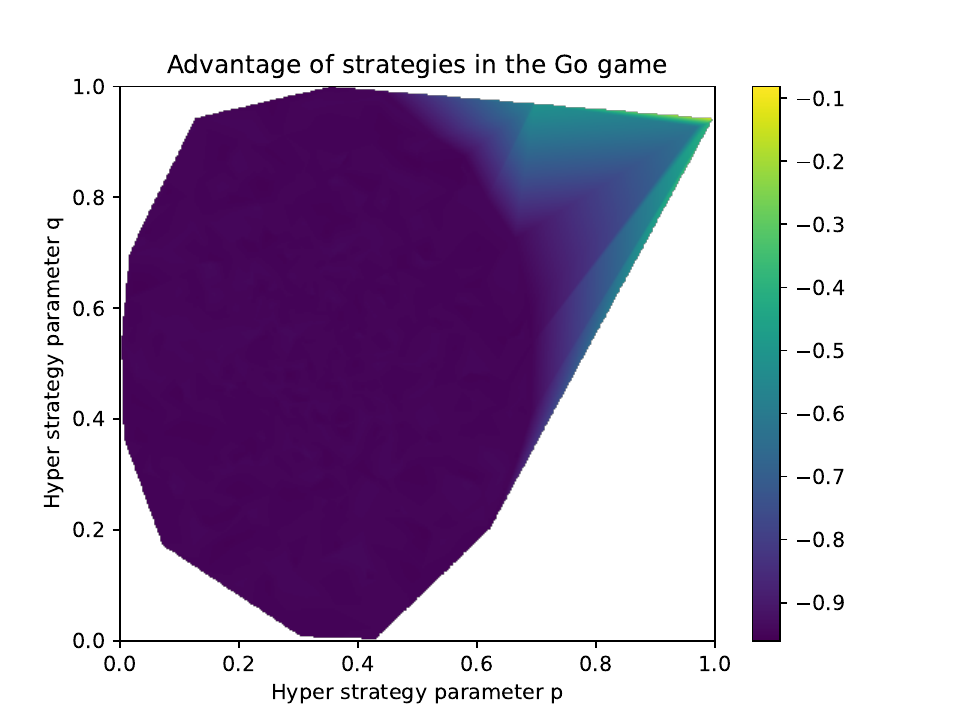}
}
\subfigure[Go (size=4)] {\label{Simplified Go game (size=4)2}
\includegraphics[width=0.22\columnwidth]{Ex_2_4.pdf}
}
\caption{The advantage distribution of strategies in various zero-sum games. Lighter colored regions indicate strategies with higher advantage.}
\label{supply2}
\end{figure}

\begin{table}[h]
\centering
\begin{tabular}{lll}
Settings               & Value           & Description                                        \\ \hline
nb\_iters              & 200             & Training iterations                                \\
meta\_solver           & fictitious play & Metasolver method                                  \\
meta\_iter             & 1000            & Iterations for Metasolver                          \\
improvement\_threshold $c_m$ & 0.03            & Convergence criteria                               \\
learning\_rate         & 0.5             & Default learning rate                              \\
num\_learners          & 4               & Number of strategies updated in each iteration     \\
num\_repeats           & 10               & Number of repetitions per experiment               \\
$l_r$              & 0.5             &      Default step size                                \\
$\lambda_d$                     & 0.5             & Diversity weight                                   \\
  \hline
\end{tabular}
\caption{Parameter setting for experiments in Zero-sum games.}
\label{table1}
\end{table}

In our setup, each experiment is repeated 10 times and the results are averaged for plotting. Within each experiment, the population is initialized randomly, and the meta-game is solved in the same manner for different algorithms. The number of learners for all algorithms, except the classic PSRO, is set to 4, which implies that four strategies within the population will be updated in each iteration. In order to more accurately compare the effiency of different algorithms in learning the strategies, we gradually increase the iterations for meta-solver. The initial iterations for meta-solver is set as 1000. Every 20 steps of training, we increase the iterations for meta-solver by 500.

Our experiments for symmetric zero-sum games are conducted in the environments used in the previous papers about PSRO algorithms. Detailed description of these game environments can be found in \cite{5,3}. Taking the AlphaStar as an example, it is derived from the experimental environment StarCraft, which is commonly used in multiagent reinforcement learning. By extracting meta-strategies in large scale extend-form game StarCraft, we can obtain a symmetric normal-form game AlphaStar. In detail, AlphaStar is a symmetric zero-sum games with dimension $888 \times 888$. Other normal-form game environments are similarly obtained by extracting meta-strategies for real world games (Go, Kuhn Poker, etc.).

Additional experiment results are shown in Figure \ref{supply1}. According to the previous work \cite{5}, the following games (\ref{Elo game},\ref{Transitive game},\ref{Triangular game},\ref{Random game of skill},\ref{Elo game + noise=0.1},\ref{3-move parity game}) has strong transitive structures. From Figure \ref{supply1}, we can see that in these games, adopting only the LA (lookahead) module with the objective of maximizing the advantage function is effective to reduce the exploitability. In those games with cyclic structures, it is necessary to adopt the diversity module in learning Nash equilibrium. We can see that A-PSRO combining LA and Diversity Module achieves the optimal results across all environments. In the stochastic game Disc game \ref{Disc game} with almost no transitive dimension, all algorithms obtain the same convergence results.

Figure \ref{supply2} shows the advantage distribution of these games. From Figure \ref{supply2}, we can see that although there may be large differences in the payoff matrices between different games, they may have similar advantage distribution. Games with the same advantage distribution have similar convergence processes of strategies when applying the PSRO algorithms to solve them, e.g. \ref{Elo game2}, \ref{Random game of skill2}, \ref{3-move parity game2}. Although the advantage function does not fully characterize the transitive dimension in zero-sum games, we believe that it has similarities to the geometric visualization of the transitive and cyclic dimensions in the previous work \cite{5}.

\subsection{A-PSRO for Solving Two-Player General-Sum Games}
\label{A-PSRO for Solving Two-Player General-Sum Games2}

\begin{table}[h]
\centering
\begin{tabular}{lll}
Settings               & Value           & Description                                        \\ \hline
nb\_iters              & 100             & Training iterations                                \\
meta\_solver           & fictitious play & Metasolver method                                  \\
meta\_iter             & 1000            & Iterations for Metasolver                          \\
num\_oracle\_repeats $k$             & 10           & Repetitions of the inner loop for strategy exploration                          \\
distribution\_type             & normal            & Gaussian distribution                    \\
distribution\_mean             & 0            & Mean value of the distribution                \\
distribution\_var             & 20            & Variance of the distribution                  \\
improvement\_threshold & 0.03            & Convergence criteria                               \\
learning\_rate         & 0.5             & Default learning rate                              \\
num\_learners          & 4               & Number of strategies updated in each iteration     \\
num\_repeats           & 100               & Number of repetitions per experiment               \\  \hline
\end{tabular}
\caption{Parameter setting for experiments in General-sum games.}
\label{table2}
\end{table}

The parameter setting of general-sum games is given in \ref{table2}. The hardware and system setup used for the experiments are the same as those for zero-sum games. Similar to zero-sum games, the initial iterations for meta-solver is set as 1000. Every 20 steps of training, we increase the iterations for meta-solver by 500.

The StagHunt game is a commomly used general-sum game environment to test the ability of algorithms to learn the optimal Nash equilibrium \cite{40}. The payoff matrix of the traditional StagHunt game is given in Table \ref{table3}. In the Stag Hunt game, both (U,L) and (D,R) are Nash equilibriums. In order to achieve a higher reward joint strategy, cooperation is required in the learning process of agents.

\begin{table}[h]
\centering
\begin{tabular}{|c|c|c|}
\hline
  & L   & R   \\ \hline
U & 30,30 & -10,-10 \\ \hline
D & -10,-10 & 20,20 \\ \hline
\end{tabular}
\caption{Traditional StagHunt game.}
\label{table3}
\end{table}

\begin{table}[h]
\centering
\begin{tabular}{|c|c|c|c|c|c|c|}
\hline
\multicolumn{1}{|c|}{}   & \multicolumn{1}{c|}{$a_1$} & \multicolumn{1}{c|}{$a_2$} & $\cdots$ & $a_i$ & $\cdots$ & $a_n$ \\ \hline
$a_1$                       & $U_1$                      & -$u$                      & -$u$ & -$u$ & -$u$ & -$u$ \\ \hline
\multicolumn{1}{|c|}{$a_2$} & \multicolumn{1}{c|}{-$u$} & \multicolumn{1}{c|}{$u$}  & $\cdots$ & -$u$ & $\cdots$ & $u$  \\ \hline
\multicolumn{1}{|c|}{$\vdots$} & \multicolumn{1}{c|}{-$u$} & \multicolumn{1}{c|}{$\cdots$} & $\ddots$ & -$u$ & $\ddots$ & $\vdots$ \\ \hline
$a_i$                       & -$u$                      & -$u$                      & -$u$ & $U_i$ & -$u$ & -$u$ \\ \hline
$\vdots$                       & -$u$                      & $\vdots$                      & $\ddots$ & -$u$ & $\ddots$ & $\vdots$ \\ \hline
$a_n$                       & -$u$                      & $u$                       & $\cdots$ & -$u$ & $\cdots$ & $u$  \\ \hline
\end{tabular}
\label{table4}
\caption{Advanced StagHunt game.}
\end{table}

In this paper, we extend the traditional StagHunt game structure to large scale general-sum game. The payoff matrix of Advanced-Staghunt is given in Table \ref{table4}. In Advanced-Staghunt, each agent has a pure strategy space $\{a_1 \cdots, a_n\}$, where $\{a_1,a_i,\cdots\}$ corresponds to the Nash equilibrium strategies resulting from cooperation. In Table \ref{table4}, $U_i$ denotes the reward corresponding to the cooperative strategy in StagHunt, which is drawn from a unifrom ditribution $\mathbf{U}[1,2]$. In order to judge whether A-PSRO deterministically converges to the optimal Nash equilibrium, we set one of those $U_i$ to 2. 

For the rest of the payoff matrix for the Advanced-StagHunt, we use the uniform distribution $u=\mathbf{U}[0,0.8]$ to fill the rewards corresponding to each joint strategy. This suggests that there are many inefficient Nash equilibria in the Advanced-StagHunt game besides the cooperative equilibrium in the shape of ($a_i,a_i$). 

In our experiments, we set $n=100$ and there are 5 cooperation equilibrium with reward $U_i \sim \mathbf{U}[1,2]$. Each PSRO algorithm was run 10 times repeatedly to solve the game and the results were averaged for presentation.

From the experiment result \ref{fg4_1}, we can see that most PSRO algorithms in the Advanced-StagHunt stagnate in the inefficient Nash equilibria. This is because the space of strategies whose strategy gradient points to a cooperative Nash equilibrium is a small proportion of the full space. In order for the agent to learn the optimal Nash equilibrium strategy, it is necessary to design reward-related objective for the agent. We can see that A-PSRO based on the advantage function effectively learns the optimal Nash equilibrium strategy, which indicates that the strategy exploration objective given in Equation (11) can improve rewards when solving general-sum games.

The optimal Nash equilibrium in the Advanced-StagHunt game is a pure strategy equilibrium. In order to test the effectiveness of A-PSRO in games where the optimal equilibrium is a mixed strategy equilibrium, we design a large-scale general-sum game Advanced-RSP with the structure similar to the traditional game Rock-Paper-Scissors. We first design the structure of the RSP in general-sum game $U_{RSP}$. The payoff matrix of $U_{RSP}$ is given in Table \ref{table5}. In $U_{RSP}$, $\epsilon$ is a random number satisfying the uniform distribution $\mathbf{U}[0,100]$. This general-sum game has the similar mixed Nash equilibrium to the traditional RSP.

Based on the $U_{RSP}$, we design the large scale general-sum game Advanced-RSP. The payoff matrix of Advanced-RSP is given in \ref{table6}. For the rest of the payoff matrix for the Advanced-RSP, we use the uniform distribution $u=\mathbf{U}[0,100]$ to fill the rewards corresponding to each joint strategy. We can easily find that each subgame corresponding to the joint strategy $(R_i,S_i,P_i)$ is a mixed Nash equilibrium. There are also other equilibra with inefficient rewards.

In our experiment, we set $n=1000$ and $i=10$. Each PSRO algorithm was run 10 times repeatedly to solve the game and the results were averaged for presentation. The experiment result is shown in Figure \ref{fg4_2}. From the figure, we can see that A-PSRO learns the optimal mixed equilibrium strategy, while all other PSRO algorithms stagnate in the inefficient Nash equilibrium.  

\begin{table}[h]
\centering
\begin{tabular}{|c|c|c|c|}
\hline
  & R     & S     & P     \\ \hline
R & 100    & 180+$\epsilon$ & $\epsilon$     \\ \hline
S & $\epsilon$     & 100    & 180+$\epsilon$ \\ \hline
P & 180+$\epsilon$ & $\epsilon$     & 100    \\ \hline
\end{tabular}
\caption{General-sum RSP structure $U_{RSP}$.}
\label{table5}
\end{table}

\begin{table}[h]
\centering
\begin{tabular}{|c|c|c|ccc|c|c|c|cll|l|l|}
\hline
                         & $a_1$                      & $\cdots$                      & \multicolumn{1}{c|}{$R_1$} & \multicolumn{1}{c|}{$S_1$} & $P_1$ & $\cdots$ & $a_j$ & $\cdots$ & \multicolumn{1}{c|}{$R_i$} & \multicolumn{1}{l|}{$S_i$} & $P_i$                      & $\cdots$                      & $a_n$                      \\ \hline
$a_1$                       & $u$                       & $\cdots$                      & \multicolumn{1}{c|}{-$u$} & \multicolumn{1}{c|}{-$u$} & -$u$ & $\cdots$ & $u$  & $\cdots$ & \multicolumn{1}{c|}{-$u$} & \multicolumn{1}{l|}{-$u$} & -$u$                      & $\cdots$                      & $u$                       \\ \hline
$\vdots$                       & $\vdots$                      & $\ddots$                      & \multicolumn{1}{c|}{-$u$} & \multicolumn{1}{c|}{-$u$} & -$u$ & $\ddots$ & $\vdots$ & $\ddots$ & \multicolumn{1}{c|}{-$u$} & \multicolumn{1}{l|}{-$u$} & -$u$                      & $\ddots$                      & $\vdots$                      \\ \hline
$R_1$                       & -$u$                      & -$u$                      & \multicolumn{3}{c|}{\multirow{3}{*}{$U_{RSP}$}}             & -$u$ & -$u$ & -$u$ & \multicolumn{1}{c|}{-$u$} & \multicolumn{1}{l|}{-$u$} & -$u$                      & -$u$                      & -$u$                      \\ \cline{1-3} \cline{7-14} 
$S_1$                       & -$u$                      & -$u$                      & \multicolumn{3}{c|}{}                                  & -$u$ & -$u$ & -$u$ & \multicolumn{1}{c|}{-$u$} & \multicolumn{1}{l|}{-$u$} & -$u$                      & -$u$                      & -$u$                      \\ \cline{1-3} \cline{7-14} 
$P_1$                       & -$u$                      & -$u$                      & \multicolumn{3}{c|}{}                                  & -$u$ & -$u$ & -$u$ & \multicolumn{1}{c|}{-$u$} & \multicolumn{1}{l|}{-$u$} & -$u$                      & -$u$                      & -$u$                      \\ \hline
$\vdots$                       & $\vdots$                      & $\ddots$                      & \multicolumn{1}{c|}{-$u$} & \multicolumn{1}{c|}{-$u$} & -$u$ & $\ddots$ & $\vdots$ & $\ddots$ & \multicolumn{1}{c|}{-$u$} & \multicolumn{1}{l|}{-$u$} & -$u$                      & $\ddots$                      & $\vdots$                     \\ \hline
$a_j$                       & $u$                       & $\cdots$                      & \multicolumn{1}{c|}{-$u$} & \multicolumn{1}{c|}{-$u$} & -$u$ & $\cdots$ & $u$  & $\cdots$ & \multicolumn{1}{c|}{-$u$} & \multicolumn{1}{c|}{-$u$} & \multicolumn{1}{c|}{-$u$} & \multicolumn{1}{c|}{$\cdots$} & \multicolumn{1}{c|}{$u$}  \\ \hline
$\vdots$                      & $\vdots$                      & $\ddots$                      & \multicolumn{1}{c|}{-$u$} & \multicolumn{1}{c|}{-$u$} & -$u$ & $\ddots$ & $\vdots$ & $\ddots$ & \multicolumn{1}{c|}{-$u$} & \multicolumn{1}{c|}{-$u$} & \multicolumn{1}{c|}{-$u$} & \multicolumn{1}{c|}{$\ddots$} & \multicolumn{1}{c|}{$\vdots$} \\ \hline
$R_i$                       & -$u$                      & -$u$                      & \multicolumn{1}{c|}{-$u$} & \multicolumn{1}{c|}{-$u$} & -$u$ & -$u$ & -$u$ & -$u$ & \multicolumn{3}{c|}{\multirow{3}{*}{$U_{RSP}$}}                                  & \multicolumn{1}{c|}{-$u$} & \multicolumn{1}{c|}{-$u$} \\ \cline{1-9} \cline{13-14} 
\multicolumn{1}{|l|}{$S_i$} & \multicolumn{1}{l|}{-$u$} & \multicolumn{1}{l|}{-$u$} & \multicolumn{1}{l|}{-$u$} & \multicolumn{1}{c|}{-$u$} & -$u$ & -$u$ & -$u$ & -$u$ & \multicolumn{3}{c|}{}                                                       & \multicolumn{1}{c|}{-$u$} & \multicolumn{1}{c|}{-$u$} \\ \cline{1-9} \cline{13-14} 
\multicolumn{1}{|l|}{$P_i$} & \multicolumn{1}{l|}{-$u$} & \multicolumn{1}{l|}{-$u$} & \multicolumn{1}{l|}{-$u$} & \multicolumn{1}{c|}{-$u$} & -$u$ & -$u$ & -$u$ & -$u$ & \multicolumn{3}{c|}{}                                                       & \multicolumn{1}{c|}{-$u$} & \multicolumn{1}{c|}{-$u$} \\ \hline
\multicolumn{1}{|l|}{$\vdots$} & \multicolumn{1}{l|}{$\vdots$} & \multicolumn{1}{l|}{$\ddots$} & \multicolumn{1}{l|}{-$u$} & \multicolumn{1}{c|}{-$u$} & -$u$ & $\ddots$ & $\vdots$ & $\ddots$ & \multicolumn{1}{c|}{-$u$} & \multicolumn{1}{c|}{-$u$} & \multicolumn{1}{c|}{-$u$} & \multicolumn{1}{c|}{$\ddots$} & \multicolumn{1}{c|}{$\vdots$} \\ \hline
\multicolumn{1}{|l|}{$a_n$} & \multicolumn{1}{l|}{$u$}  & \multicolumn{1}{l|}{$\cdots$} & \multicolumn{1}{l|}{-$u$} & \multicolumn{1}{c|}{-$u$} & -$u$ & $\cdots$ & $u$  & $\cdots$ & \multicolumn{1}{c|}{-$u$} & \multicolumn{1}{c|}{-$u$} & \multicolumn{1}{c|}{-$u$} & \multicolumn{1}{c|}{$\cdots$} & \multicolumn{1}{c|}{$u$}  \\ \hline
\end{tabular}
\label{table6}
\caption{Advanced RSP game.}
\end{table}

We also perform experiments in randomly generated games that feature the same reward distribution. The dimension of random generated games are $1000 \times 1000$. In these games, each element of the payment matrix is generated by a normal distribution with mean $\mu=0$ and variance $\sigma^2=20$. To test the convergence results of different algorithms in random generated games, we let the PSRO algorithms operate in 100 independently generated game environments and the results were averaged for presentation. The experiment result is shown in Figure \ref{fg4_3}. From Figure \ref{fg4_3}, we can see that A-PSRO achieves the optimal reward of the joint strategy.


\subsection{A-PSRO for Solving Multi-Player Games}
\label{A-PSRO for Solving Multi-Player Games2}
\begin{table}[h]
\centering
\begin{tabular}{lll}
Settings               & Value           & Description                                        \\ \hline
players              & 3             & Number of agents in the game                                \\
nb\_iters              & 50             & Training iterations                                \\
meta\_solver           & fictitious play & Metasolver method                                  \\
meta\_iter             & 10000            & Iterations for Metasolver                          \\
distribution\_type             & normal            & Gaussian distribution                    \\
distribution\_mean             & 0            & Mean value of the distribution                \\
distribution\_var             & 20            & Variance of the distribution                  \\
improvement\_threshold & 0.03            & Convergence criteria                               \\
learning\_rate         & 0.5             & Default learning rate                              \\
num\_learners          & 4               & Number of strategies updated in each iteration     \\
num\_repeats           & 4               & Number of repetitions per experiment               \\
$\lambda_d$                     & 0.5             & Diversity weight                                   \\
   \hline
\end{tabular}
\caption{Parameter setting for experiments in multi-player games.}
\label{table7}
\end{table}

The parameter setting of multi-player games is given in \ref{table7}. The hardware and system setup used for the experiments are the same as those for zero-sum games. In our multi-player game experiments, we adopt randomly generated games that feature the same reward distribution. 

In multi-player zero-sum games, we use randomly generated symmetric games with dimension $20\times20\times20$. This is because we have found in our experiments that reducing exploitability in larger-scale games requires very large computational complexity. We believe that a comparison with other algorithms in the setting of this size is sufficient to demonstrate effectiveness. During the generation of these games, we added constraints to avoid generating strong pure strategies, which substantially increased the difficulty of strategy learning. 

In the generation of multi-player general-sum games, we use the structure similar to the Advanced-Staghunt game with dimension $10\times10\times10$, and set the reward of the optimal equilibrium strategy to 90. A detailed game generation process will be published along with the code in the future.

\end{document}